\documentclass[%
 reprint,
 nosort,
 citeautoscript,
 amsmath,
 amssymb,
 showkeys,
 aps,
10pt,
]{revtex4-1}

\usepackage{graphicx}
\usepackage{dcolumn}
\usepackage{bm}
\usepackage{hyperref}
\usepackage{todonotes}
\usepackage{float}
\usepackage{mathtools}
\usepackage{csquotes}
\usepackage{cleveref}
\usepackage{braket}
\usepackage{subfig}
\usepackage{graphicx}
\usepackage{mathptmx}
\usepackage{caption}
\usepackage[utf8]{inputenc}
\usepackage{bbding}
\usepackage[normalem]{ulem}

\usepackage{pdfpages} 
\usepackage{pgffor} 
\usepackage{xr} 
\usepackage{siunitx}

\def\supplementfilename{SI}

\pdfximage{\supplementfilename.pdf}
\def\numbersupplementpages{\the\pdflastximagepages}

\newif\ifarXiv
\arXivtrue 

\makeatletter
\AtBeginDocument{\let\LS@rot\@undefined}
\makeatother

\setcitestyle{super}

\begin{document}

\title{Revealing molecule-internal mechanisms that control phonon heat transport \\ through single-molecule junctions by a genetic algorithm} 

\author{Matthias Blaschke}
\affiliation{Institute of Physics and Center for Advanced Analytics and Predictive Sciences, University of Augsburg, 86135 Augsburg, Germany}
\author{Fabian Pauly}
\affiliation{Institute of Physics and Center for Advanced Analytics and Predictive Sciences, University of Augsburg, 86135 Augsburg, Germany}

\date{\today}

\begin{abstract}

Measurements of the thermal conductance of single-molecule junctions have recently been reported for the first time. It is presently unclear, how much the heat transport can be controlled through molecule-internal effects. 
The search for molecules with lowest and highest thermal conductance is complicated by the gigantic chemical space. 
Here we describe a systematic search for molecules with a low or a high phononic thermal conductance using a genetic algorithm. Beyond individual structures of well performing molecules, delivered by the genetic algorithm, we analyze patterns and identify the different physical and chemical mechanisms to suppress or enhance phonon heat flow. In detail, mechanisms revealed to reduce phonon transport are related to the choice of terminal linker blocks, substituents and corresponding mass disorder or destructive interference, meta couplings and molecule-internal twist. For a high thermal conductance, the molecules should instead be rather uniform and chain-like. The identified mechanisms are systematically analyzed at different levels of theory, and their significance is classified. Our findings are expected to be important for the emerging field of molecular phononics.
\end{abstract}

\maketitle

Electronic properties of single-molecule junctions can nowadays be characterized rather routinely \cite{ratner2013brief}. Different transport phenomena such as rectification \cite{loertscher2012transport, elbing2005single}, switching \cite{liao2010cyclic,irie2002digital}, and thermoelectric energy conversion \cite{gemma2023full, reddy2007thermoelectricity, cui2018peltier} have been studied at the molecular scale so far. Quantum interference effects and sharp transmission resonances, originating from molecular orbitals, allow for a precise control of electronic characteristics \cite{Lambert:ChemRev2015}, even enabling quantum distance sensing \cite{reznikova2021substitution, stefani2018large,Schosser:Nanoscale2022,Hsu:ChemSci2022,li2021mechanical}. Such prototypical single-molecule studies yield fundamental insights of how important quantum mechanical coherence is at a given temperature.

Beyond charge transport, heat transport constitutes a field in its own right. It is well known that the electrical conductivity of common bulk materials at room temperature varies over more than 20 orders of magnitude \cite{haynes2016crc}. The thermal conductivity originates in contrast not only from electronic carriers, but lattice vibrations, mass transport or radiative effects may add. Through the Wiedemann-Franz law the electronic contribution to the thermal conductivity is proportional to the electrical conductivity, and a similar variation of the electronic thermal conductivity can thus be expected. But the additional heat transport contributions substantially reduce the variability of the thermal conductivity of common materials to only some 6 orders of magnitude at room temperature \cite{MajumdarLimit}, by preventing it to vanish. For bulk materials without voids, electronic and lattice vibrational parts of the thermal conductivity are most important.

Measurements of the thermal conductance of single-molecule junctions have only recently become possible \cite{cui2019thermal, mosso:NanoLett2019}. For this reason, thermal transport and its control in molecular junctions are experimentally largely unexplored.
In single-molecule junctions, electronic, phononic and radiative effects are expected to be relevant for the  thermal conductance \cite{KloecknerThCondMerit}. But only the electronic and phononic contributions are related to the molecular structure, since radiative heat transport is mainly determined by the electrode geometry \cite{KloecknerThCondMerit}. Assuming that the connection between electrical conductance and electronic thermal conductance through the Wiedemann-Franz law remains valid at the atomic scale \cite{Cui:Science2017,Kloeckner:PRB2017-WF,Buerkle:NanoLett2018}, the lattice contribution needs to be well characterized. It is also the most relevant part for electrically rather insulating molecules \cite{cui2019thermal}. Here we explore the phononic or lattice vibrational contribution to the thermal conductance for single-molecule junctions. We search especially for molecules with a minimal or maximal thermal conductance. 

The chemical space allows for an almost infinite number of molecular structures, complicating a systematic search for molecules with low or high thermal conductance. For this purpose, we use a genetic algorithm to optimize the molecules \cite{genetic_algorithm}. In this way we avoid time-consuming trial and error cycles, which are often unsuccessful. Genetic algorithms are a powerful optimization method, and they are used, e.g., for minimum-energy structure
prediction at the nanoscale \cite{lazauskas2017efficient, sierka2010synergy}, drug design or de novo discovery of molecules \cite{spiegel2020autogrow4, sousa2021combining, ghaheri2015applications, wang2024genetic,rasmussen2023toward}. Compared to other machine learning approaches, it is appealing that genetic algorithms are not a black box method. We are not only interested in the best individuals, but in common features of these candidates. Unexpected solutions can often be seen in the results of genetic algorithms and machine learning approaches \cite{lohn2004evolutionary, bentley2002creative, melnikov2018active, PhysRevA.107.010101}. In fact, through the genetic algorithm and its best individuals we identify four different mechanisms to suppress the phononic thermal conductance: presence of (i) appropriate building blocks at the molecular ends, (ii)  substituents, (iii) meta coupling, and (iv) nonvanishing molecule-internal twist. Uniform chains instead yield the largest phononic thermal conductance. 

Although there are several theoretical studies in the literature aimed at tailoring the phononic thermal conductance of single-molecule junctions, the understanding of mechanisms controlling it is still limited. For example, destructive interferences \cite{klockner2017tuning, markussen2013phonon, sangtarash2020radical, famili2017suppression} and heteroatoms \cite{sadeghi2019quantum,D0TA07095E,PhysRevE.94.052123} have been proposed as ways to suppress the thermal conductance. Molecule-internal twist angles were considered \cite{sergueev2011efficiency, sadeghi2019quantum}, but the effects were studied at discrete angles or intermixed with substituents effects. Also different electrode-molecule anchor groups were analyzed \cite{noori2021effect}. Overall, the effects have been investigated in a rather isolated fashion, and the magnitude of changes observed was significantly smaller compared to possible variations in electronic transport. 
This calls for an investigation in a unified framework that allows to reveal the most relevant degrees of freedom, as we present it here.

\section*{\label{sec:theoretical_approach}Theoretical approach}
\subsection*{Genetic encoding and genetic algorithm}
\begin{figure}
    \includegraphics[width=1.0\columnwidth]{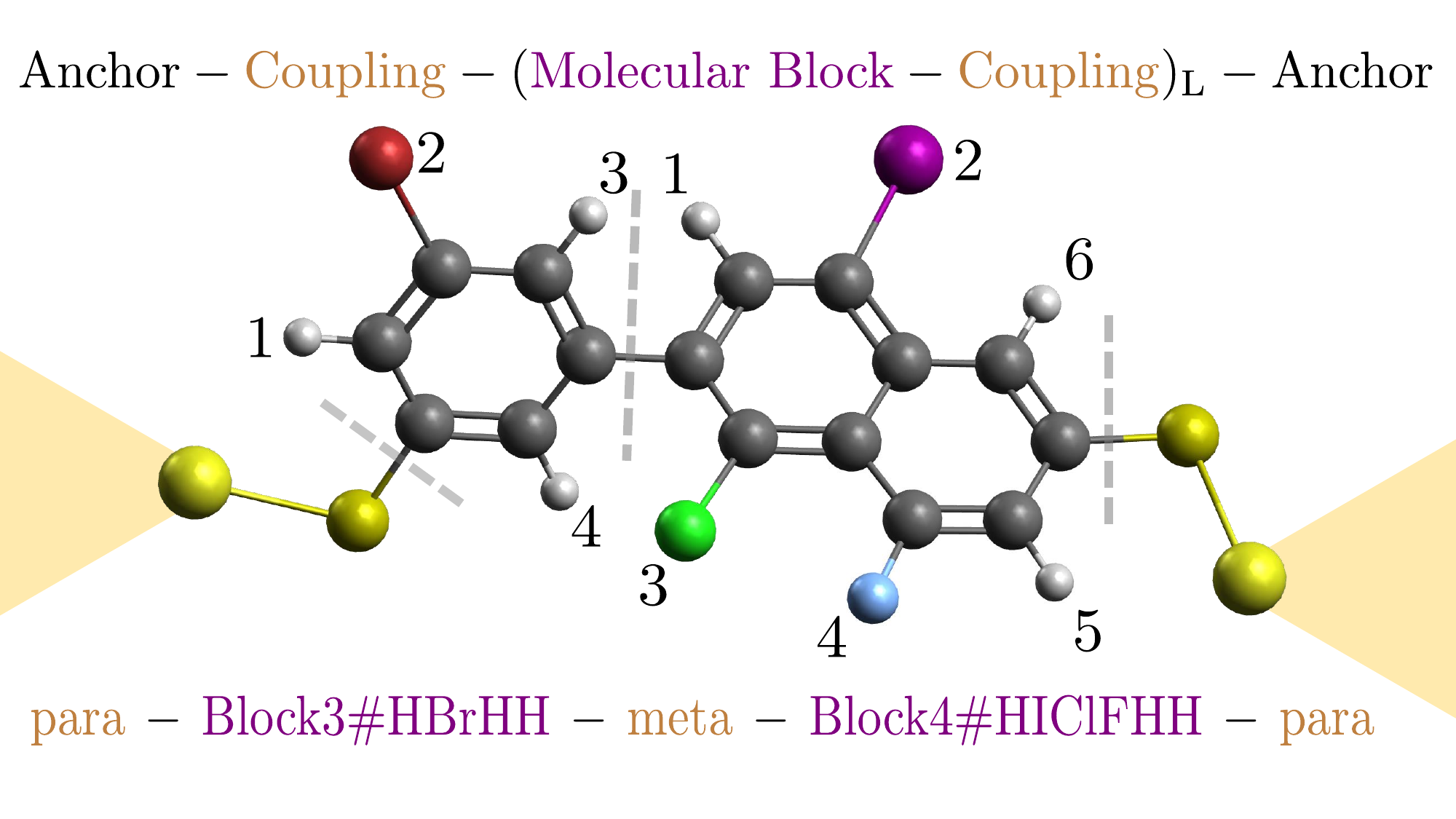}
    \caption{Genetic encoding of molecules. The general scheme is shown in the top row. The encoding consists of molecular blocks, derived from pristine building blocks by attaching substituents, and couplings.
    The molecule, depicted in the middle, is encoded according to this scheme in the bottom row. Dashed gray lines mark the block limits. Throughout this study the substituents are shown in the following colors: fluorine, cyan; chlorine, green; bromine, red; iodine, purple. Substituents may be attached at the positions that are indicated by black numbers for each block. Hydrogen atoms or the corresponding substituents are specified after a '\#' token in the encoding in the numbered order. Anchor groups, i.e.\ a sulfur atom (yellow) on the left and right side of the molecule, are omitted from the encoding, since they are always the same and thus redundant. Gold tips of the envisioned metal-molecule-metal junction are represented by yellow triangles. Depending on the simulation, we may add a single gold atom (yellow, larger diameter than sulfur atoms), as shown here, or saturate the sulfur anchors with a single hydrogen at each side.}
    \label{fig:encoding}
\end{figure}

By representing the molecular structure as genetic information, the genetic encoding determines the chemical search space accessible for the discovery of optimal individuals. Its choice is hence a crucial design step. Machine learning techniques increasingly become part of the chemical sciences \cite{meyers2021novo,Generative_Models}, and the representation of molecules by now is virtually a research field of its own \cite{wigh2022review, KRENN2022100588}. Our study concentrates on molecules that are contacted by two metallic electrodes to form metal-molecule-metal junctions, where common representations are not applicable. Therefore, we developed a tailored genetic encoding, which is designed for the representation of molecules exhibiting well-defined electrode contact points and which incorporates the important degrees of freedom controlling phononic heat transport. In the top row of figure~\ref{fig:encoding} the general structure of our encoding scheme is specified, below which the encoding string of an example molecule is shown.

\begin{figure}
    \includegraphics[width=1.0\columnwidth]{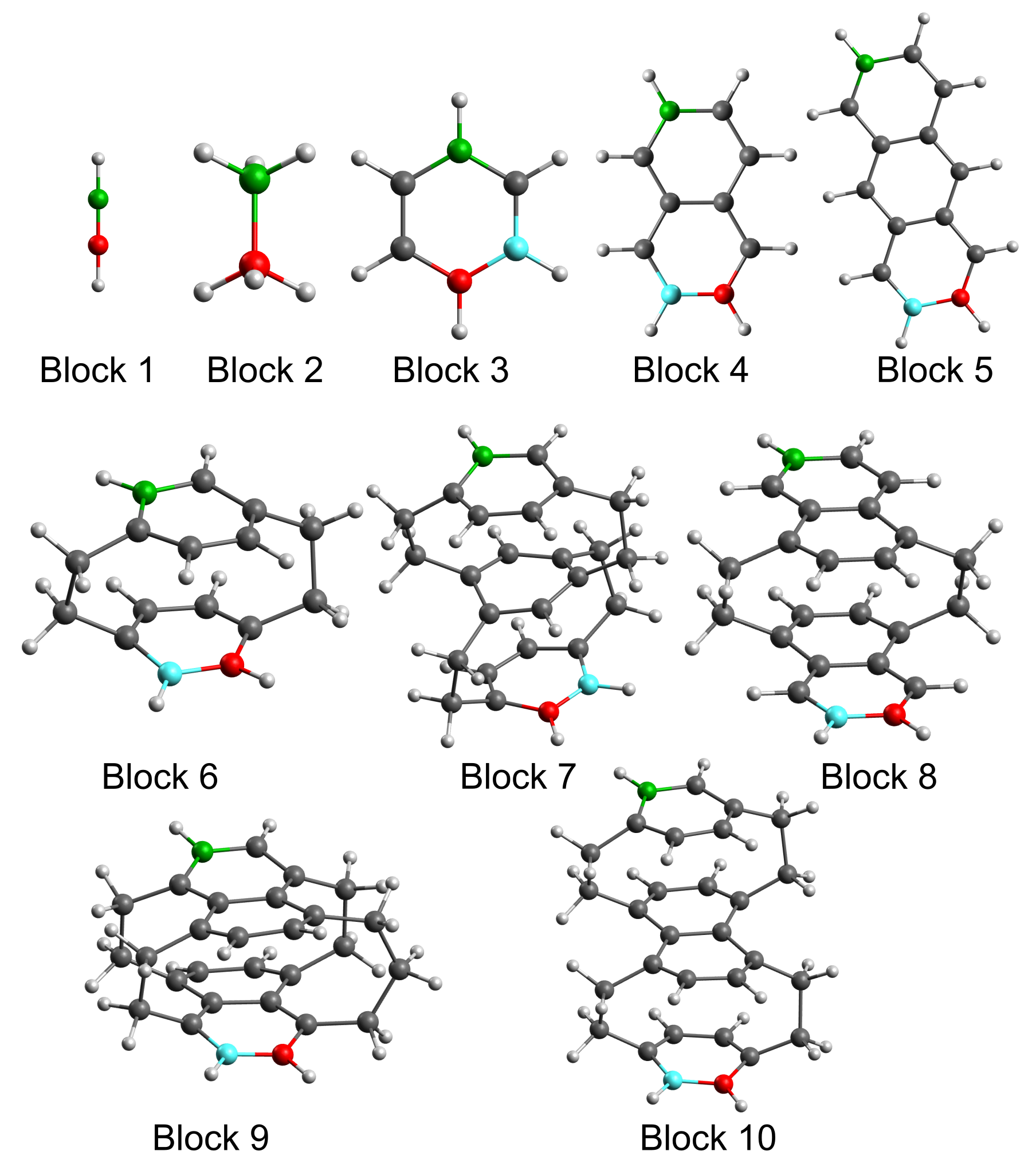}
    \caption[Molecular building blocks]{Pristine molecular building blocks used to construct molecules. Green-colored carbon atoms indicate the "left" coupling point. Red carbon atoms mark the para-coupling position on the "right", whereas cyan carbon atoms mark the meta-coupling position. Blocks with only red carbon atoms, see blocks 1 and 2, do not differ in para and meta connection. For the sake of clarity the multiplicity of bonds is not shown.}
    \label{fig:molecular_blocks}
\end{figure}

Our goal is to cover a large part of the chemical space investigated theoretically so far for tailored single-molecule thermal conductance \cite{klockner2017tuning,chen2020local, markussen2013phonon}. For this purpose we have significantly advanced the genetic encoding, initially developed in Ref.~\citenum{genetic_algorithm} for the search of mechanosensitive molecules, 
to identify the most important mechanisms controlling heat transport. Focusing on hydrocarbon-based structures, the main idea is to construct molecules from predefined building blocks and to link a certain number of them to assemble the full molecule. An overview of the pristine building blocks is given in figure~\ref{fig:molecular_blocks}. By varying the number and kind of building blocks, their type of linking and substituent attachments, crucial degrees of freedom such as the molecular length, para vs.\ meta coupling and mass disorder are introduced.

In the set of pristine building blocks, see figure~\ref{fig:molecular_blocks}, we consider  benzene, naphthalene and anthracene (blocks 3 to 5), because these conjugated components have often been used to investigate phonon transport phenomena in molecular junctions \cite{klockner2017tuning, ramezani2021side, noori2021effect, ramezani2020enhanced, zeng2020nanoscale}. We have additionally included some $\pi$-stacked components \cite{wu2008molecular} (blocks 6 to 10), based on benzenes or naphthalenes and linked covalently by ethyl bridges, since $\pi$-$\pi$ interaction may be a strategy to suppress the thermal conductance \cite{li2017strategy, kirvsanskas2014designing, burkle2015first}. To build even simpler molecules such as alkanes \cite{klockner2016length,meier2014length}, we added ethane (block 2). Acetylene (block 1) can be used to assemble linear chains or molecules like OPE3, which represent other prototypical model systems \cite{markussen2013phonon,klockner2017tuning}. 

Since side groups or substituents have been reported to cause interesting transport effects such as destructive interferences \cite{klockner2017tuning, markussen2011graphical, burkle2015first, klockner2016length, markussen2013phonon, ramezani2021side, zeng2020nanoscale}, we added a new degree of freedom to the encoding \cite{genetic_algorithm}. We allow every hydrogen atom except for those removed for linking of different building blocks (green, red or cyan carbon atoms in figure~\ref{fig:molecular_blocks}) and hydrogen atoms in ethyl brides of $\pi$-stacked blocks (blocks 6 to 10 in figure~\ref{fig:molecular_blocks}) to be replaced by the halogens fluorine, chlorine, bromine or iodine. We restrict ourselves to these single-atom substituents to avoid additional vibrational modes inside the side-groups, which would complicate the analysis of phonon transport, and yet to offer a wide range of different atomic masses. An example of the encoding, including substituents, is shown in the bottom row of figure~\ref{fig:encoding}. The substituents are specified for every block, separated by the character '\#'. The characters following '\#' denote the atoms in the order of the numbered positions. These positions are fixed for each pristine building block in the database, so that the derived building blocks containing the substituents are precisely defined. 

Anchoring groups connect a molecule to macroscopic electrodes. They are added at the beginning and at the end of the molecule. This allows for the construction of chemically valid structures with well defined junction geometry. 
The anchor groups at the molecule-electrode interface can influence phononic transport \cite{noori2021effect} and could be included in the encoding. We restrict ourselves however to thiol anchors, since we are primarily interested in robust molecule-internal features determining heat transport. Furthermore, we focus on gold electrodes since this noble metal has been used in most measurements of the single-molecule thermal conductance so far \cite{cui2019thermal, mosso:NanoLett2019}, and sulfur anchors couple well to gold. We include up to one gold atom of the electrodes on each side of the molecule in the extended central cluster \cite{pauly2008cluster} (see the methods part \ref{sec:fitness_calc} for further details). If no gold atom is attached, sulfur is saturated with a hydrogen atom. Since the sulfur atoms at the beginning and at the end of the molecule are redundant, we do not specify them in the encoding string, see figure~\ref{fig:encoding}. 

Optimization using the genetic algorithm emulates an evolution by means of selection, crossover and mutation for several generations. Each generation consists of a certain number of individuals. After a random initialization the so-called fitness value is calculated for each individual. The fitness calculation is fundamental to the performance of the genetic algorithm as it assesses to what extent the molecules behave as desired. To determine molecules with low phononic thermal conductance, we choose the fitness function basically inversely proportional to the phonon thermal conductance as
\begin{equation}
    f = \frac{1}{\kappa_\mathrm{ph}(600~\mathrm{K})/\kappa_0+c_\mathrm{SP}\sum_ic_i+c_\mathrm{SA}\alpha}.
    \label{eq:fitness_low-kappa}
\end{equation}
To optimize for high phononic thermal conductance instead, we set the fitness function proportional to the heat conductance:
\begin{equation}
    f=\kappa_{\mathrm{ph}}(600~\text{K})/\kappa_0, \label{eq:fitness_high-kappa}
\end{equation}
Here, $\kappa_0=1~\mathrm{pW/K}$ is a normalization. We evaluate the temperature-dependent thermal conductance $\kappa_{\mathrm{ph}}(T)$ at $T=600$~K, because it is well saturated at this temperature for gold electrodes. 
In the evolution loop, we determine molecular geometries and force constants, characterized by the dynamical matrix $\bm{D}$ and needed for the calculation of $\kappa_{\mathrm{ph}}(T)$, using xTB \cite{grimme2017robust, bannwarth2019gfn2} in the GFN1  parameterization \cite{grimme2017robust}.

We determine the elastic and phase-coherent phonon thermal conductance $\kappa_\text{ph}(T)$ in the fitness functions, equations~(\ref{eq:fitness_low-kappa}) and (\ref{eq:fitness_high-kappa}), through the Landauer-B\"uttiker formalism, as described in detail in the methods part~\ref{sec:fitness_calc}. Note that we exclusively focus on phonons and neglect possible electronic contributions to heat transport. This approximation may not be well justified for very short molecules\cite{cui2019thermal}, but for long molecules the off-resonant electrical conductance decays exponentially and with it the electronic heat conductance based on the Wiedemann-Franz law \cite{klockner2016length,Cui:Science2017,Kloeckner:PRB2017-WF,cui2019thermal}. 
Since the Landauer-Büttiker formalism neglects inelastic phonon-phonon and electron-phonon interactions, the molecules should conversely not be too long. The inelastic phonon mean free path in bulk gold at room temperature has been reported to be between 1 to 10~nm \cite{McGaughey_meanfree}, and we expect that molecules should not be much longer than that. (In the following, the lengths of molecules studied will be around 1~nm.) However, the precise length may depend sensitively on the molecules studied and the environmental conditions. The static junction geometries explored here as well as the assumption of elastic phase-coherent transport are certainly best fulfilled at low temperatures\cite{mingo2006anharmonic,Mueller_PhysRevB, volz_anharmonic, wang2008quantum}.

Apart from the thermal conductance, we may steer additional molecular properties through the fitness function. 
First, the number of substituents should not be too large in order to obtain simple structures. We take this into account by introducing an optional substituent penalty (SP) and choose a reciprocal relationship between fitness and the SP. Each substituent in the encoding of a molecular structure can be penalized with a specific cost $c_i$ for the given atom type $i=\text{F}, \text{Cl}, \text{Br}, \text{I}$, as indicated in expression~\eqref{eq:fitness_low-kappa}. Naturally, no penalty is imposed for hydrogen atoms, and $c_\text{H}=0$. The factor $c_\mathrm{SP}$ weights the overall importance of the SP inside the fitness function. Second, it is often challenging to synthesize molecules suggested by artificial intelligence methods \cite{gao2020synthesizability}. For this reason, we consider the synthetic accessibility (SA) score $\alpha$ in equation~\eqref{eq:fitness_low-kappa}. It is described in Ref.~\citenum{ertl2009estimation}, and we evaluate it through RDKit. It estimates the SA of drug-like molecules and assigns values from  1, for easy to make, to 10, for very difficult to make. To determine $\alpha$ for the xTB-relaxed structure of each candidate, we replace  terminal gold atoms by hydrogen to obtain thiol termini. We include the SA score reciprocally in the fitness, since a lower $\alpha$ should be rewarded, while a difficult synthesis should be penalized. We weight the SA score inside $f$ by $c_\mathrm{SA}$.  
In the different evolution runs, to be discussed in the results section, we will vary $c_\mathrm{SP}$ and $c_\mathrm{SA}$, see table~\ref{tab:run_configuration}. The SP is only used for optimization to low thermal conductance, see equation~\eqref{eq:fitness_low-kappa}, since mass disorder is disadvantageous for high phonon transmission. Since structures exhibiting a high $\kappa_\text{ph}(T)$ turn out to be linear chains, we do not need to consider the SA in equation~\eqref{eq:fitness_high-kappa} either.

Based on the calculated fitness $f$, the best molecular structures are chosen from a population to produce offspring and to form a new generation. In the process the genetic information of the parents is inherited by the offspring and is modified by genetic operations such as crossover and mutation. These new combinations of characteristics might be better with an increased chance of surviving due to a higher fitness value. Poorly performing individuals in contrast are penalized by a low fitness value. This process is iterated until a convergence criterion is reached or until the evolution is stopped for other reasons. 

For the analysis of transport properties we will sometimes study the propagator elements $\left[\bm{G}^\text{r}(E)\right]_{(i,\mu),(j,\nu)}$. Here, $i$ is an atomic index belonging to an atom coupled to the left electrode, and $j$ is an index of an atom coupled to the right electrode. The indices $\mu,\nu=x,y,z$ characterize the Cartesian components of displacement of atoms $i,j$, respectively. For the sake of simplicity we denote such matrix elements of the Green's function, see also equation~\eqref{eq:propagator}, as $P_{\mu\nu}(E)$. 
We take the square of the absolute value of these matrix elements $|P_{\mu\nu}(E)|^2$, in order to approximate the transmission, see equation~\eqref{eq:transmission_ph}.

We refer the reader to the methods part\ref{sec:Methods} for further information on methodological aspects such as the genetic encoding and related size of chemical space, the determination of the phonon thermal conductance, and the evolution loop. Note also that the full program code is publicly available at Zenodo \cite{Blaschke_gaPh}, and the described procedures may hence be directly examined inside the code.

\section*{\label{sec:results}Results of evolution runs}

\subsection*{Optimization for low phonon heat conductance}

\begin{figure*}[t!]
    \includegraphics[width=2.0\columnwidth]{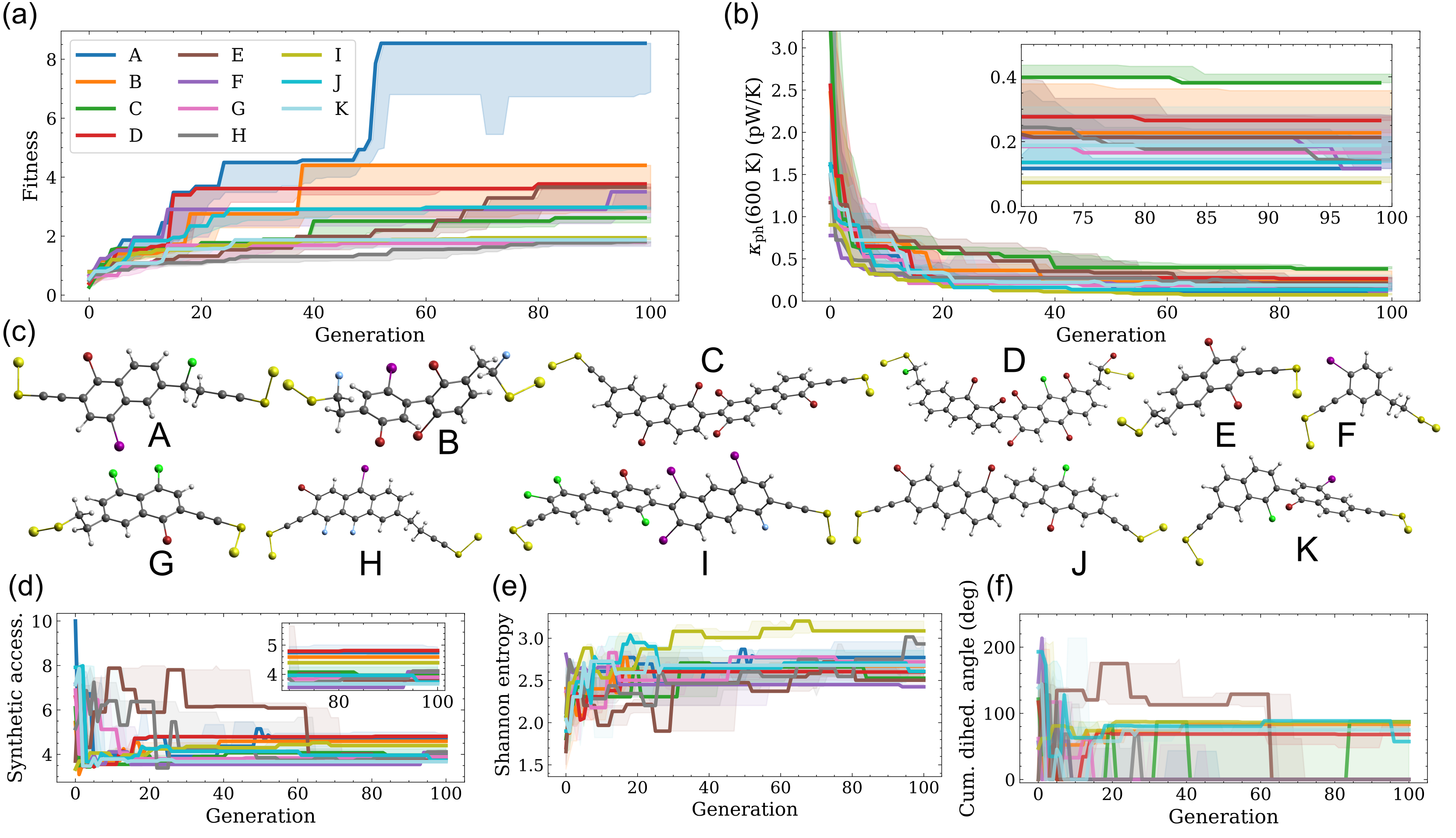}
    \caption{Results of different evolution runs optimizing molecules for low phononic thermal conductance. (a) Fitness plotted as a function of the generation for evolution runs A to K. Parameters used for the fitness function in equation~\eqref{eq:fitness_low-kappa} are listed in table~\ref{tab:run_configuration}. (b) Thermal conductance $\kappa_\mathrm{ph}(T)$ at $T=600$~K plotted against the generation number. The inset shows an enlarged section of the plot at high generations. Candidates in the initial populations show typically thermal conductance values larger than $25~\mathrm{pW/K}$. (c) Molecular structures of the fittest individuals in the last generation of each evolution run. (d) SA for each generation and evolution run. For the assessment of the SA, terminal gold atoms are removed and replaced by hydrogens, yielding isolated molecules with SH termination. The inset shows an enlarged section of the plot at high generations. (e) Shannon entropy for all evolution runs. (f) Same as (e) but for the cumulated dihedral angle. In panels (a), (b), (d), (e) and (f), solid lines show the values of the fittest individual and shaded regions around the solid lines visualize the scatter range of the four best performing candidates.}
    \label{fig:result_overview}
\end{figure*}

Figure~\ref{fig:result_overview} summarizes the results for optimization to low thermal conductance. Parameters used in the fitness function~\eqref{eq:fitness_low-kappa} are specified in table~\ref{tab:run_configuration} for each run. 

Figure~\ref{fig:result_overview}(a) shows the fitness values of the best performing molecule as a function of generation for every evolution run. The shaded regions furthermore depict the scatter range of the top four molecules, which are directly transferred to the next generation by elitism. Due to the different definitions of the fitness function in equation~\eqref{eq:fitness_low-kappa}, the fitness values of the different runs are not fully comparable. A clear increase of $f$ can be observed for all runs, which begins to saturate starting typically at generation $50$. The fitness value assesses the relative performance of molecules in each run. 

\begin{table}[t!]
\begin{tabular}{|l|l|l|l|l|}
\hline
Run &  SP \& SA weights & Substituents & \begin{tabular}[c]{@{}l@{}}Substituent\\ cost\end{tabular} & \begin{tabular}[c]{@{}l@{}} Symmetry \\ enforced\end{tabular} \\ \hline
A &  $c_\mathrm{SP}=c_\mathrm{SA}=0$ & F, Cl, Br, I & 0 & no \\ \hline
B &  $c_\mathrm{SP}=c_\mathrm{SA}=0$ & F, Cl, Br, I & 0 & yes \\ \hline
C, D &  $c_\mathrm{SP}=c_\mathrm{SA}=0$ & Cl, Br & 0 & yes \\ \hline
E &  $c_\mathrm{SP}= 1; c_\mathrm{SA}=0$ & F, Cl, Br, I & \begin{tabular}[c]{@{}l@{}} $c_\mathrm{Hydrogen}$=0 \\  $c_\mathrm{Halogen}$=0.03 \end{tabular} & no  \\ \hline
F &  $c_\mathrm{SP}= 1; c_\mathrm{SA}=0$ & F, Cl, Br, I & \begin{tabular}[c]{@{}l@{}} $c_\mathrm{Hydrogen}$=0 \\  $c_\mathrm{Halogen}$=0.1 \end{tabular} & no \\ \hline
G, H &  $c_\mathrm{SP}= 0; c_\mathrm{SA}=0.1$ & F, Cl, Br, I & 0 & no \\ \hline
I &  $c_\mathrm{SP}= 0; c_\mathrm{SA}=0.1$ & F, Cl, Br, I & 0 & yes \\ \hline
J &  $c_\mathrm{SP}= 1; c_\mathrm{SA}=0$ & F, Cl, Br, I & \begin{tabular}[c]{@{}l@{}} $c_\mathrm{Hydrogen}$=0 \\  $c_\mathrm{Halogen}$=0.03 \end{tabular} & yes \\ \hline
K &  $c_\mathrm{SP}= 1; c_\mathrm{SA}=0$ & F, Cl, Br, I & \begin{tabular}[c]{@{}l@{}} $c_\mathrm{Hydrogen}$=0 \\  $c_\mathrm{Halogen}$=0.1 \end{tabular} & yes \\ \hline
\end{tabular}
\caption{Definition of the fitness function~\eqref{eq:fitness_low-kappa} for the evolution runs presented in figure~\ref{fig:result_overview}. Label of the evolution run, weights for SP and SA, allowed substituents, penalty per substituent $c_i$, and option for enforcing symmetry. In all runs, the generation limit is $100$ and the population size $100$. The length of each candidate is limited between $L_\text{min}=2$ and $L_\text{max}=4$ building blocks. The probability for each mutation operation is set to $p_\text{m}=0.7$. The best four individuals are transferred to the next generation by the elitism step, see figure~\ref{fig:evoloop}. The selection of molecules for the mating pool uses the $k$-tournament method with $k=20$. In the random generation of individuals and in the substituent mutation operation, selection of hydrogen is $18$ times more likely than those of halogens.}
\label{tab:run_configuration}
\end{table}

Our primary target is the phononic thermal conductance, whose evolution is plotted in figure~\ref{fig:result_overview}(b). All evolution runs exhibit a fast decrease of the thermal conductance at the beginning of the evolution process. Similar to the trend of the fitness values, a saturation is typically observed after generation 50. Lowest phononic thermal conductances turn out to be on the order of $0.1$~pW/K.

Let us now discuss, how different fitness functions influence 
the molecular structures of the best performing candidates in the last generation, which are depicted in figure \ref{fig:result_overview}(c). 
All molecules show similar chemical and structural characteristics that will be analyzed systematically later. The fitness of the molecule that reaches the highest value of $f$ in figure~\ref{fig:result_overview}(a) belongs to run A and is fully determined by the thermal conductance ($c_\mathrm{SP}=c_\mathrm{SA}=0$), all substituents are allowed and no symmetry is enforced. By symmetry we mean that the sequence of pristine building blocks may optionally be symmetrized with respect to the center of the molecule. Details are given in the methods part\ref{sec:evolution_loop}.  Despite possessing the highest fitness value, the resulting structure of run A does not exhibit the lowest thermal conductance. A high SA score, see figure~\ref{fig:result_overview}(d), furthermore indicates that it is not easy to produce. To restrict chemical space, we enforce symmetry regarding the building blocks for evolution B. Since the substituents and couplings are not symmetrized, the chemical structure is still somewhat asymmetric 
and features a rather complex chemical structure with different types of halogen substituents. For this reason, we have restricted the selection of substituents to chlorine and bromine for evolution runs C and D and again enforce symmetry with respect to the pristine building blocks. The complexity of the molecular structure for run C decreases, as indicated by the low SA score in figure~\ref{fig:result_overview}(d). However, the chemical structure for run D remains complicated. As a further step we introduce a penalty for each substituent in runs E and F by setting $c_\mathrm{SP}>0$ in equation~\eqref{eq:fitness_low-kappa}. In this way a structure with many substituents is deprecated. Indeed, this procedure results in molecules that have fewer substituents, lower chemical complexity, and yet show a significantly suppressed thermal conductance. 
For cases G, H, I the SA is directly incorporated into the fitness function by setting $c_\mathrm{SA}>0$, see table~\ref{tab:run_configuration}. To avoid getting trapped in a specific region of chemical space at the beginning of the evolution, we use a rather low weight $c_\mathrm{SA}=0.1$. Symmetry is enforced only for run I. Although the lowest overall phonon thermal conductance is achieved for run I, the evolutions G to I do not significantly improve the results in terms of chemical synthesizability of the structures, as compared to the previous ones. Furthermore, it should be noted that the SA in figure~\ref{fig:result_overview}(d) evolves to values around 4 for almost all evolution runs, even if it is not explicitly included in the fitness function. Molecules in this range are somewhat less accessible than catalog molecules \cite{ertl2009estimation}. The last evolution runs J and K correspond to E and F, where we penalize the substituents. In addition, the symmetry is now enforced for run J and K. In this way, the lowest SAs are achieved, see figure~\ref{fig:result_overview}(d). 

In summary, the genetic algorithm robustly optimizes molecules for low phononic heat conductance with all presented specifications of the fitness function. Consideration of additional properties through appropriate constraints can enhance the quality of the candidates with regard to these aspects and simplify their molecular structures. 

In the following, we do not want to focus on specific optimal molecules, however, but rather work out common characteristics. In this way, we will learn more about physical principles that are crucial for the design of molecules with low heat conductance. All optimized molecules with strongly suppressed thermal conductance, displayed in figure~\ref{fig:result_overview}(c), 
are indeed characterized by the following features: (i) acetylene or ethyl terminal blocks, (ii) substituents, (iii) one meta coupling, (iv) internal torsion induced by substituents.

The length of the optimized molecules does not show a special behavior. All runs (except for E, F, G) result in structures with the maximum number of four allowed building blocks. It was shown that the thermal transport properties of rather short and electrically insulating molecules only moderately depend on molecular length \cite{klockner2016length,cui2019thermal}. Therefore, we will not analyze the length dependence of transport in more detail. We note that our theory applies in the phase-coherent elastic transport regime, which should be realized for molecules shorter than the inelastic scattering length.

Only blocks $1$ to $5$ from figure~\ref{fig:molecular_blocks} are observed in the final structures of figure~\ref{fig:result_overview}(c). The $\pi$-stacked blocks $6$ to $10$ do not appear to be advantageous to realize a low phononic thermal conductance. 

Substituents can suppress phonon transport \cite{klockner2017tuning, markussen2013phonon,sadeghi2019quantum,D0TA07095E,PhysRevE.94.052123, bro2024heavy} by introducing mass disorder or destructive quantum interferences. 
The Shannon entropy \cite{shannon1948mathematical} of the corresponding genetic information can quantify this disorder. We define the Shannon entropy by $S=-\sum_i p_i \log(p_i)$, where the summation runs over all characters in the encoding string and $p_i$ is the probability of a character to be sampled from the set of available characters in the string. It can be interpreted as the lower bound of information needed to encode the molecule \cite{cover1999elements}. 
Interestingly, the Shannon entropy in figure~\ref{fig:result_overview}(e) increases during the evolution. This provides evidence that the disorder itself contributes to the suppression of the thermal conductance.

We find two distinct classes of molecules in figure~\ref{fig:result_overview}(c): twisted or planar structures. To study this, figure~\ref{fig:result_overview}(f) shows the cumulated dihedral angle of the candidates. We determine this angle by separating the relaxed structure into its building blocks. In the next step a plane is fitted through all atoms of each building block, containing the center of mass. The dihedral angle between neighboring building blocks is determined through the angle between the normal vectors of the fitted planes. Since for acetylene and ethyl end groups (blocks 1 and 2) planes are not well defined, the corresponding dihedral angle with a neighboring block is set to $0^\circ$. The cumulated dihedral angle is determined by adding the absolute values of all dihedral angles between neighboring blocks and hence distinguishes twisted from planar molecules. Evolution runs A, E, F, G, H yield planar structures. The remaining runs show total internal torsional angles of $90^\circ$ with a small spread. The torsion is typically induced by steric repulsion at block-block linkages due to halogen substituents. The twisted and planar molecular classes basically separate within the first $20$ generations. It should be noted that the molecule-internal torsion is expected to depend on temperature. 
Strong steric repulsions may however lead to high rotational barriers, yielding stable conformers \cite{Pauly:PRB2008}.

Another common feature is that all the optimized molecules possess acetylene or ethyl groups that are connected to the sulfur anchor on the left and right side. This property thus appears to be crucial for suppressing phonon heat transport. Finally, all molecules except for run A in figure~\ref{fig:result_overview}(c) exhibit exactly one meta coupling, whereas the remaining couplings are in a para configuration. 

\begin{figure}[]
    \includegraphics[width=1.0\columnwidth]{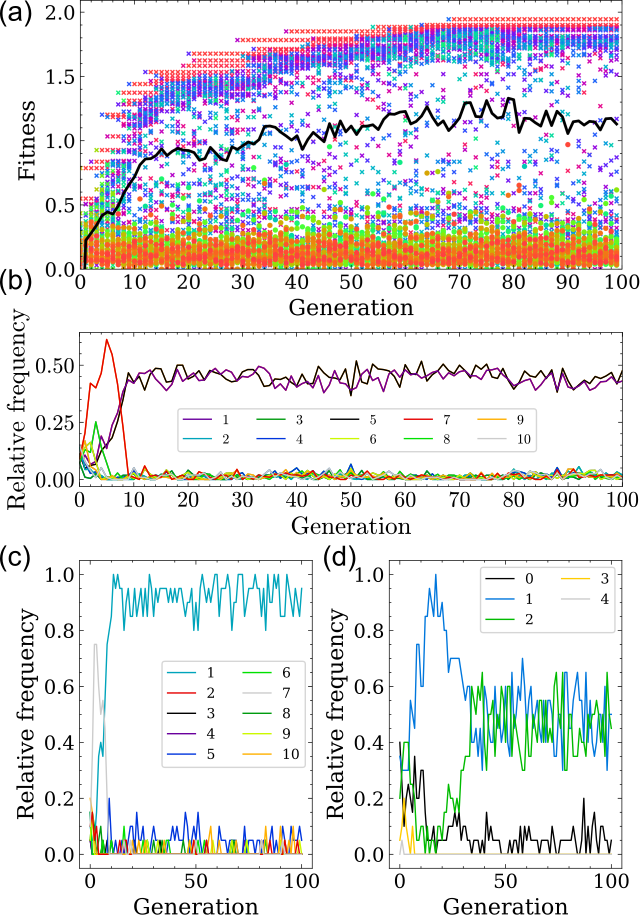}
    \caption{Analysis of the evolution run I in figure~\ref{fig:result_overview}, leading to the lowest overall heat conductance. (a) Fitness values of the whole population as a function of generation. Crosses mark molecular candidates evolved through selection, crossover and mutation, dots indicate randomly generated individuals. The black solid line shows the mean fitness. Relative frequency of (b) all molecular building blocks and (c) building blocks in the first and last positions of the encoding string for each generation. (d) Statistics of coupling classes for each generation, defined by counting the number of meta couplings in each molecular structure. Remaining couplings are in para configuration. Panel (b) shows the statistics of the whole generation, in panels (c) and (d) the statistics consider the top $20$ individuals of each generation. For panels (b) and (c), block numbers in the legend correspond to those in figure~\ref{fig:molecular_blocks}.}
    \label{fig:run_37_ana}
\end{figure}

In the next section, the common features (i) to (iv) will be analyzed in detail. Let us now illustrate, how the evolutionary processes work at the example of run I, which leads to the lowest $\kappa_\text{ph}(T)$ in figure~\ref{fig:result_overview}. The evolution of fitness values with generation number is depicted in figure~\ref{fig:run_37_ana}(a).

The building block distribution for the whole generation is depicted in figure~\ref{fig:run_37_ana}(b) during the evolution. After the random distribution of blocks in generation $1$, block $7$ dominates up to generation $8$. Around generation $10$ the frequencies of blocks $1$ and $5$ rise and block $7$ disappears. The frequencies of blocks $1$ and $5$ saturate around generation $15$, and the block statistics remains stable up to the generation limit. The comparison to figure~\ref{fig:run_37_ana}(a) shows that the fitness values keep increasing even though the statistics of pristine blocks remains unchanged. This demonstrates that reconfiguration of the substituents plays an important role during the evolution. 

Figure~\ref{fig:run_37_ana}(c) shows the statistics of the terminal building blocks of the 20 best individuals. Due to the symmetrization, first and last blocks (i.e.\ those blocks on the left and right that are connected to sulfur anchor atoms) are identical. In the first generation all blocks are distributed uniformly. Subsequently the frequencies of blocks $1$ and $7$ increase. However, block $7$ becomes extinct around generation $10$ and block $1$ prevails.

Finally, couplings are analyzed in figure~\ref{fig:run_37_ana}(d). Our classification is based on the number of meta couplings that occur inside a molecule. We show the characteristics of the best $20$ candidates for each generation. In the beginning the classes 0 to 4 are distributed nearly uniformly. In generation $25$ the highest frequency is observed for the class with exactly one meta coupling, whereas the other classes almost vanish at this point. Near generation $30$ a strong increase of the class with two meta couplings sets in, and the classes with one or two meta couplings saturate around $50\%$ relative frequency for higher generations. In the end the best performing molecule exhibits only one meta coupling.

\subsection*{Optimization for high phonon heat conductance}

\begin{figure}[]
    \includegraphics[width=1.0\columnwidth]{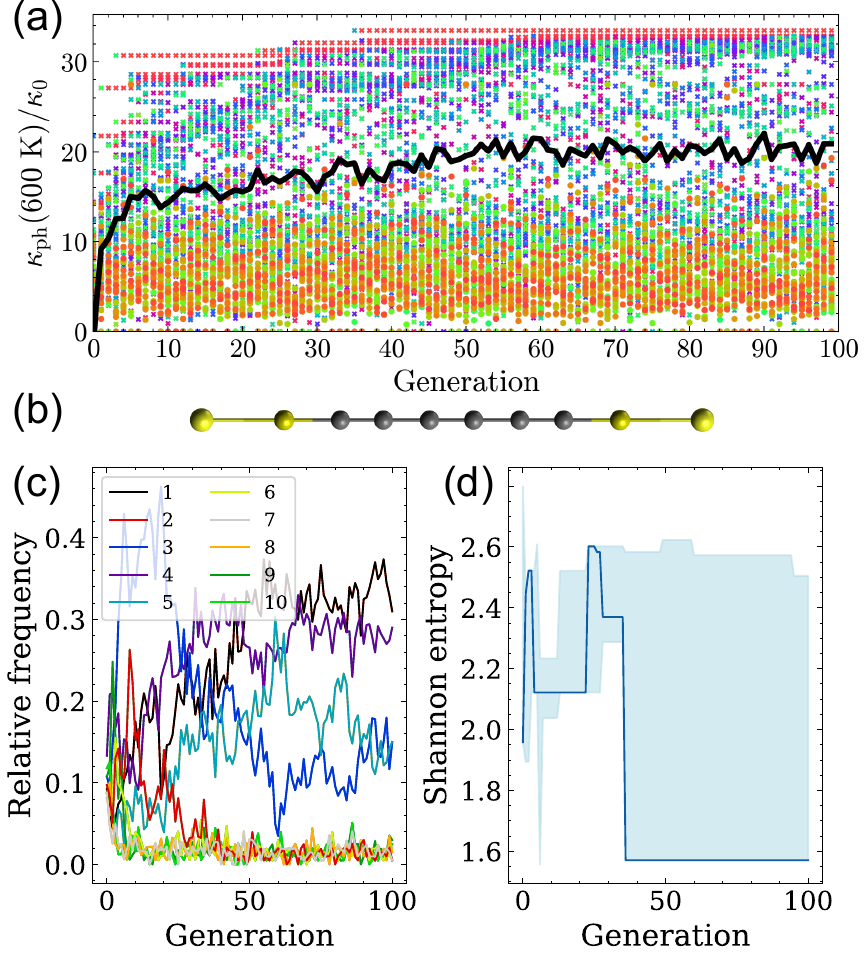}
    \caption{Optimization of molecules for high phonon thermal conductance. (a) Fitness values as a function of generation. Crosses mark molecular candidates evolved through selection, crossover and mutation. Dots indicate randomly generated individuals. The black line shows the mean fitness value. According to equation~\eqref{eq:fitness_high-kappa} the fitness is directly proportional to the thermal conductance $\kappa_\mathrm{ph}(600~\mathrm{K})$, measured in units of $\kappa_0=1$~pW/K. (b) The best performing molecule of the last generation. (c) Relative frequency of molecular blocks for each generation, considering the top $20$ individuals of each generation. Block numbers in the legend correspond to those of figure~\ref{fig:molecular_blocks}. (d) Shannon entropy of the best performing molecule with the highest thermal conductance in each generation. Shaded regions visualize the scatter range of the best four individuals.}
    \label{fig:run_35_ana}
\end{figure}

An evolution optimizing for high phonon thermal conductance is presented in figure~\ref{fig:run_35_ana}. In this case, we use the fitness function~\eqref{eq:fitness_high-kappa}, which is directly proportional to the heat conductance. 

The fitness values of the whole population in each generation are depicted in figure~\ref{fig:run_35_ana}(a). Here, the fitness values are already rather high in the first few generations, and only a moderate increase is seen until the generation limit is reached.  
Similar to the optimization for low thermal conductance, evolved individuals show superior performance to the randomly generated individuals. 

The best performing molecule is displayed in figure~\ref{fig:run_35_ana}(b) and consists of a linear chain formed by three acetylene blocks. The linear Au$_1$-S-C bonds at the termini in figure~\ref{fig:run_35_ana}(b) arise from our way of optimizing the molecular geometry, since we start from a linear configuration in the xTB relaxation. To check the robustness of the results, we examine chains with one to five acetylene building blocks and bent Au$_1$-S-C anchors or just with thiol anchors in the Supporting Information. Again, the linear chain consisting of three units of block 1 yields the highest  thermal conductance. 

How blocks are distributed in the top $20$ candidates of each generation is displayed in figure~\ref{fig:run_35_ana}(c). In contrast to figure~\ref{fig:run_37_ana}(b) more than two molecular blocks show high frequency during the evolution, namely blocks $1$, $3$, $4$ and $5$. Blocks $6$ to $10$ are revealed to be unimportant. Chains of ethyl groups might be imagined as well. However, according to figure \ref{fig:run_35_ana}(c), block $2$ does not prevail. 

The Shannon entropy of the best performing molecule of each generation is depicted in figure~\ref{fig:run_35_ana}(d). The entropy drops to lower values during the evolution and saturates around generation $35$. 
This behavior is consistent with the expectation that low atomic disorder facilitates energy transport.

\section*{\label{sec:discussion}Discussion}

In this section, we will study the identified mechanisms for suppressed thermal conductance in detail. Each mechanism is investigated separately, and we refer the interested reader to the Supporting Information for further analysis. 

\subsection*{Mechanisms to suppress the phononic thermal conductance}

\subsubsection*{Terminal building blocks}\label{sec:mech_terminal-building-blocks}

Most of the molecules with strongly suppressed phonon thermal conductance in figure \ref{fig:result_overview} are terminated either by acetylene or ethyl building blocks. The effect of anchoring groups on $\kappa_\text{ph}(T)$ was investigated in Ref.~\citenum{noori2021effect} using pyridyl, thiol, methyl sulfide and carbodithioate, and the large influence was reported to be due to the different electrode-molecule coupling strength. As the acetylene units in our study are directly connected to the sulfur atoms, these terminal building blocks can be seen as a "linker group" or "extended anchor". 

\begin{figure}
    \includegraphics[width=1.0\columnwidth]{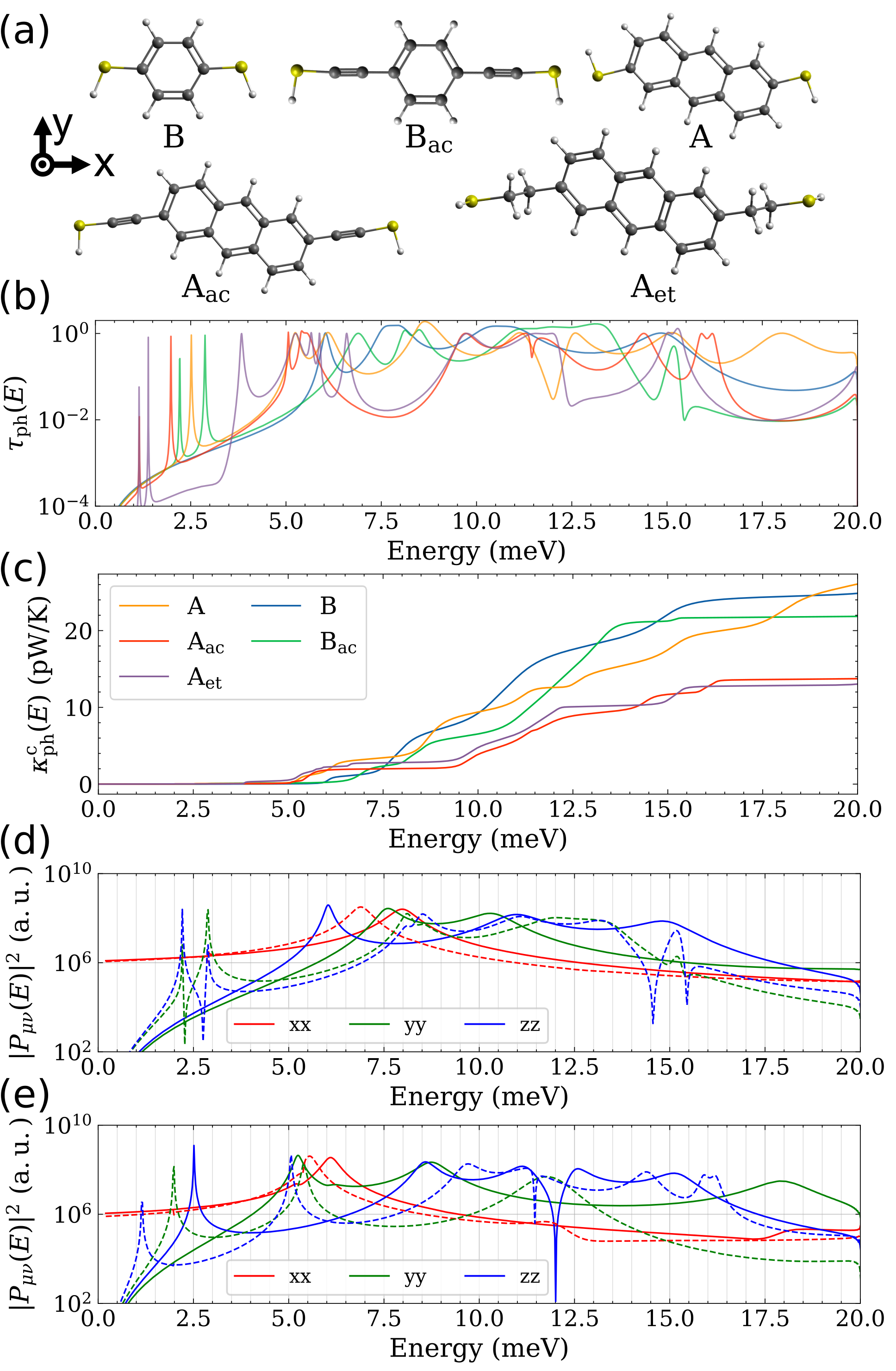}
    \caption{Phononic transport calculations for the molecules with and without terminal acetylene or ethyl units. (a) Molecular structures studied. (b) Transmission as a function of energy for the molecules depicted in (a). (c) Cumulative thermal conductance as a function of energy. Relevant absolute square of the propagator elements $|\bm{P}_{\mu\nu}(E)|^2$ for (d) benzene-derived molecules ($\mathrm{B}$, $\mathrm{B_{ac}}$) and (e) anthracene-derived molecules ($\mathrm{A}$, $\mathrm{A_{ac}}$). The pair of indices $\mu,\nu$ is shown in the legend. Solid lines are used for benzene (B) and anthracene (A), dashed ones for the structures with acetylene units. The coordinate system is aligned as indicated in (a). Mixed modes do not have a significant contribution and are therefore not shown.}
    \label{fig:triple_comp}
\end{figure}

To explore the influence of terminal building blocks on the phonon thermal conductance, we compare the transport properties of benzene and anthracene with acetylene linkers ($\mathrm{B_{ac}}$, $\mathrm{A_{ac}}$) and without ($\mathrm{B}$, $\mathrm{A}$) in figure~\ref{fig:triple_comp}. In addition, we show the results for terminal ethyl building blocks attached to anthracene ($\mathrm{A_{et}}$). The molecular structures are depicted in figure~\ref{fig:triple_comp}(a). We concentrate on thiol terminated molecules to avoid different gold-sulfur configurations after geometry optimization. Gold atoms of the Au$_1$-S group would normally not lie inside the plane of the benzene or anthracene rings \cite{Buerkle:PhysRevB2012}, complicating the interpretation of $\mu\nu=xx$, $yy$ and $zz$-modes of terminal propagator elements $P_{\mu\nu}(E)$. 

Below the studied molecular structures, we show the phonon transmissions in figure \ref{fig:triple_comp}(b). Significant differences for the two benzene and three anthracene configurations, respectively, are evident. By analyzing the cumulative thermal conductance $\kappa_\mathrm{ph}^\text{c}(E,\text{300~K})$ in figure~\ref{fig:triple_comp}(c), decisive phonon energies can be identified. Note also that $\kappa_\mathrm{ph}^\text{c}(E,\text{300~K})$ yields the thermal conductance $\kappa_\mathrm{ph}(300~\mathrm{K})$ at sufficiently high energies, when saturation sets in. The thermal conductance of $\mathrm{B_{ac}}$ is significantly lower than that of B, and the same holds true for $\mathrm{A_{ac}}$ and $\mathrm{A_{et}}$ compared to $\mathrm{A}$. 
 
Let us first analyze the benzene structures. The cumulative thermal conductance of B in figure~\ref{fig:triple_comp}(c) is larger than that of $\mathrm{B_{ac}}$ at most energies. An important exception is the region around 13.5~meV, and figure~\ref{fig:triple_comp}(d) reveals that mainly $yy$- and $zz$-modes at 13.5~meV lead to a significant contribution to the thermal conductance of $\mathrm{B_{ac}}$. The thermal conductance of B finally grows beyond that of $\mathrm{B_{ac}}$ above $15$~meV, since $zz$- and $yy$-modes are comparatively strongly suppressed for $\mathrm{B_{ac}}$. Thus the total thermal conductance of $\mathrm{B_{ac}}$ is smaller than that of B, since transversal $zz$- and $yy$-modes in $\mathrm{B_{ac}}$ are suppressed in this high energy range.

For anthracene structures, the cumulative thermal conductance of $\mathrm{A}$ lies basically above that of $\mathrm{A_{ac}}$ and $\mathrm{A_{et}}$ across the whole energy range. Particularly important for the enhanced conductance of A are the modes around $8.5$~meV, $13~\mathrm{meV}$, $15~\mathrm{meV}$ and $17~\mathrm{meV}$. As figure~\ref{fig:triple_comp}(e) shows, at $8.5$~meV they are of type $yy$ and $zz$, at $13~\mathrm{meV}$ and $15~\mathrm{meV}$ of type $zz$, and at $17~\mathrm{meV}$ of type $yy$. An additional propagator analysis, which includes the behavior for $\mathrm{A_{et}}$, is presented in the Supporting Information. In summary, the difference in the thermal conductance of the studied anthracene derivatives arises from the suppression of transversal modes with $yy$- and $zz$-character through the acetylene or ethyl end groups. Longitudinal modes of $xx$-type are largely unaffected by the linker groups.  

We explain the suppression of the thermal conductance due to terminal building blocks by a mismatch in the force constants between the acetylene or ethyl linkers and the benzene-like backbone. We model the $\mathrm{B_{ac}}$ system in the Supporting Information using a nearest-neighbor tight-binding approach, which supports this hypothesis. Overall, the terminal acetylene and ethyl blocks can be regarded as mode filters that suppress the thermal conductance. Especially in-plane transversal $yy$-modes and out-of-plane transversal \textit{zz}-modes are suppressed at high energies.

High electrical conductance combined with suppressed thermal conductance is crucial for enhanced thermoelectric efficiency \cite{cui2017perspective, gemma2021roadmap, goldsmid2013thermoelectric}. Since acetylene blocks have little effect on the electrical conductance \cite{reznikova2021substitution, stefani2018large}, they appear to be appropriate elements to design molecular junctions with improved thermoelectric performance.

\subsubsection*{Substituents}\label{sec:mech_substituents}

Mass disorder, as quantified by the Shannon entropy in figure~\ref{fig:result_overview}(e), reduces the phononic thermal conductance \cite{sadeghi2019quantum,D0TA07095E,PhysRevE.94.052123, bro2024heavy}. A more detailed view shows that substituents can induce Fano antiresonances \cite{klockner2017tuning, markussen2013phonon, sangtarash2020radical, famili2017suppression}, also called destructive interferences. Depending on the mass and structure of the side groups, the antiresonances lie at different phonon energies. 
Those dips are truly molecular features, which are independent of the embedding self-energy \cite{markussen2013phonon}. 

For illustration, we present a detailed analysis for molecule F from figure \ref{fig:result_overview}. The molecule exhibits two additional elements that reduce the thermal conductance, namely a meta coupling and appropriate terminal building blocks. Thus, the transport should already be suppressed. We will now analyze the influence of the iodine substituent by comparing the structure $\mathrm{F}$ with the same structure $\mathrm{\widetilde{F}}$, where the iodine substituent is replaced by a hydrogen atom. 

\begin{figure}
    \includegraphics[width=1.0\columnwidth]{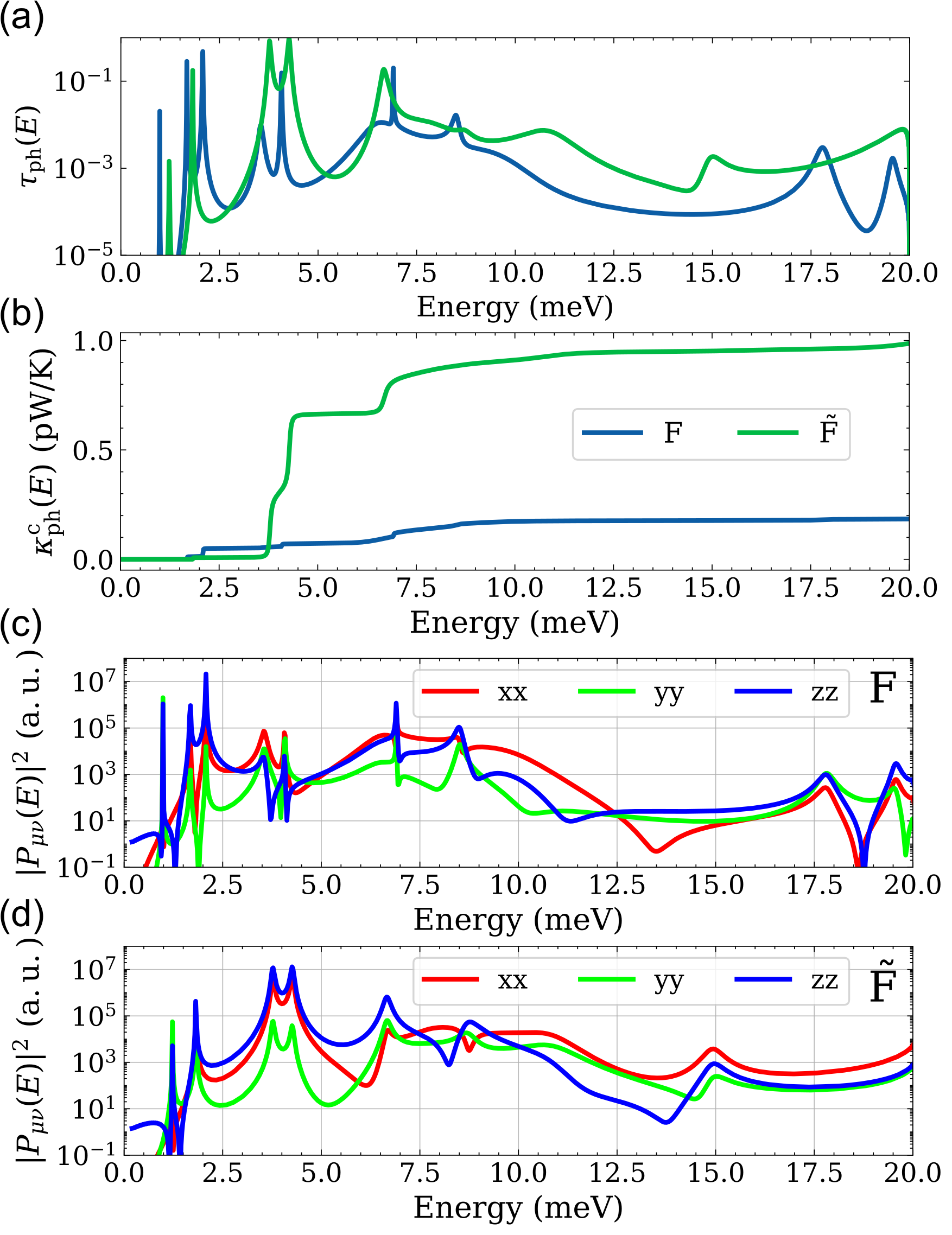}
    \caption{(a) Transmission as a function of energy for structure F from figure \ref{fig:result_overview} and structure $\mathrm{\widetilde{F}}$, where the iodine substituent of F has been replaced with hydrogen. (b) Cumulative thermal conductance at 300~K as a function of energy. Diagonal components of terminal propagator elements for (c) $\mathrm{F}$ and (d) $\tilde{\mathrm{F}}$ as a function of energy.}
    \label{fig:run25_ana}
\end{figure}

The phonon transmission curves, calculated for $\mathrm{F}$ and $\mathrm{\widetilde{F}}$, are depicted in figure \ref{fig:run25_ana}(a). The most significant differences are located at around 4~meV and 6.5~meV. At 4~meV, structure $\mathrm{\widetilde{F}}$ shows two peaks with perfect transmission reaching 1, whereas $\mathrm{F}$ exhibits smaller narrow transmission resonances in this energy region. Near 6.5~meV, the situation is similar with a broader transmission resonance for $\widetilde{\text{F}}$ but a much narrower one for F. The cumulative thermal conductance in figure~\ref{fig:run25_ana}(b) confirms that the identified energy ranges around 4~meV and 6.5~meV are indeed responsible for the main differences in the transport properties of $\mathrm{F}$ and $\mathrm{\widetilde{F}}$. 
The terminal propagator elements for $\mathrm{F}$ and $\mathrm{\widetilde{F}}$, respectively, are analyzed in figure~\ref{fig:run25_ana}(c) and \ref{fig:run25_ana}(d). The substituted structure F exhibits a typical Fano-shaped feature around 4~meV and 6.5~meV, whereas structure $\mathrm{\widetilde{F}}$ shows pronounced peaks, resembling the transmission resonances. This is particularly well visible in the \textit{zz}-component. Furthermore, in the propagator elements a pronounced destructive interference effect is visible for F at around $18.5~\mathrm{meV}$ in $xx$ and $zz$ components. 
However, the modes of $yy$ character remain active, and the transmission of the reference structure $\widetilde{\text{F}}$ shows no major contribution to the thermal conductance at this energy either. The destructive interference near 18.5~meV is therefore not significant in comparison. 

In summary, the iodine substituent induces clear destructive interferences. If the destructive interference occurs in an energy range, where the unsubstituted molecular backbone has maximum transmission, the thermal conductance is severely reduced. 
This analysis shows that the genetic algorithm places substituents in such a way that the energetic position of a resulting destructive interference is optimized.

\subsubsection*{Meta coupling}\label{sec:mech_meta-coupling}

Similar to substituents, meta couplings can induce destructive interferences and thus suppress the thermal conductance \cite{klockner2017tuning,markussen2013phonon}. Since the influence of meta couplings has already been studied extensively in the theoretical literature \cite{markussen2013phonon,markussen2011graphical}, we do not provide an analysis here. The effect is shortly discussed in the Supporting Information.

\subsubsection*{Twist angle}\label{sec:mech_twist-angle}

The dependence of the electrical conductance on twist angle is theoretically and experimentally well understood for $\pi$-conjugated molecules \cite{Pauly:PRB2008,Venkataraman:Nature2006,mishchenko2010influence}. In the experimental studies, the twist angle of a biphenyl was varied gradually and locked using different side groups.  Similar effects were already investigated for phononic heat transport \cite{sergueev2011efficiency, sadeghi2019quantum}, but the influence of the attached side groups and the twist angle has not been separated. Since the side groups do not participate strongly in electronic transport, they can be neglected there to a large extent. In contrast, phononic transport is strongly influenced by side groups and resulting interferences \cite{klockner2017tuning, famili2017suppression}, see the discussion of figure~\ref{fig:run25_ana}\ref{sec:mech_substituents}. Therefore, these properties need further analysis. 

Similar to the electronic case and following the results of the genetic algorithm in figure~\ref{fig:result_overview}, we lock the dihedral angle between the rings of a biphenyl (BP) molecule using halogen substituents. To focus the analysis on the twist angle, we artificially set the masses of all substituents to the hydrogen mass when computing the phononic thermal conductance, but we also present the thermal conductance values without this mass manipulation, see table~\ref{tab:kappa_BP}. 
In line with the studies on terminal building blocks, we reduce the anchoring group at each side to a single sulfur atom saturated with hydrogen. We proceed like this in order to avoid incomparable gold-sulfur configurations after geometry optimization and to facilitate the interpretation of the character of vibrational modes in terms of Cartesian components of propagator elements. 

\begin{figure}[t]
    \includegraphics[width=1.0\columnwidth]{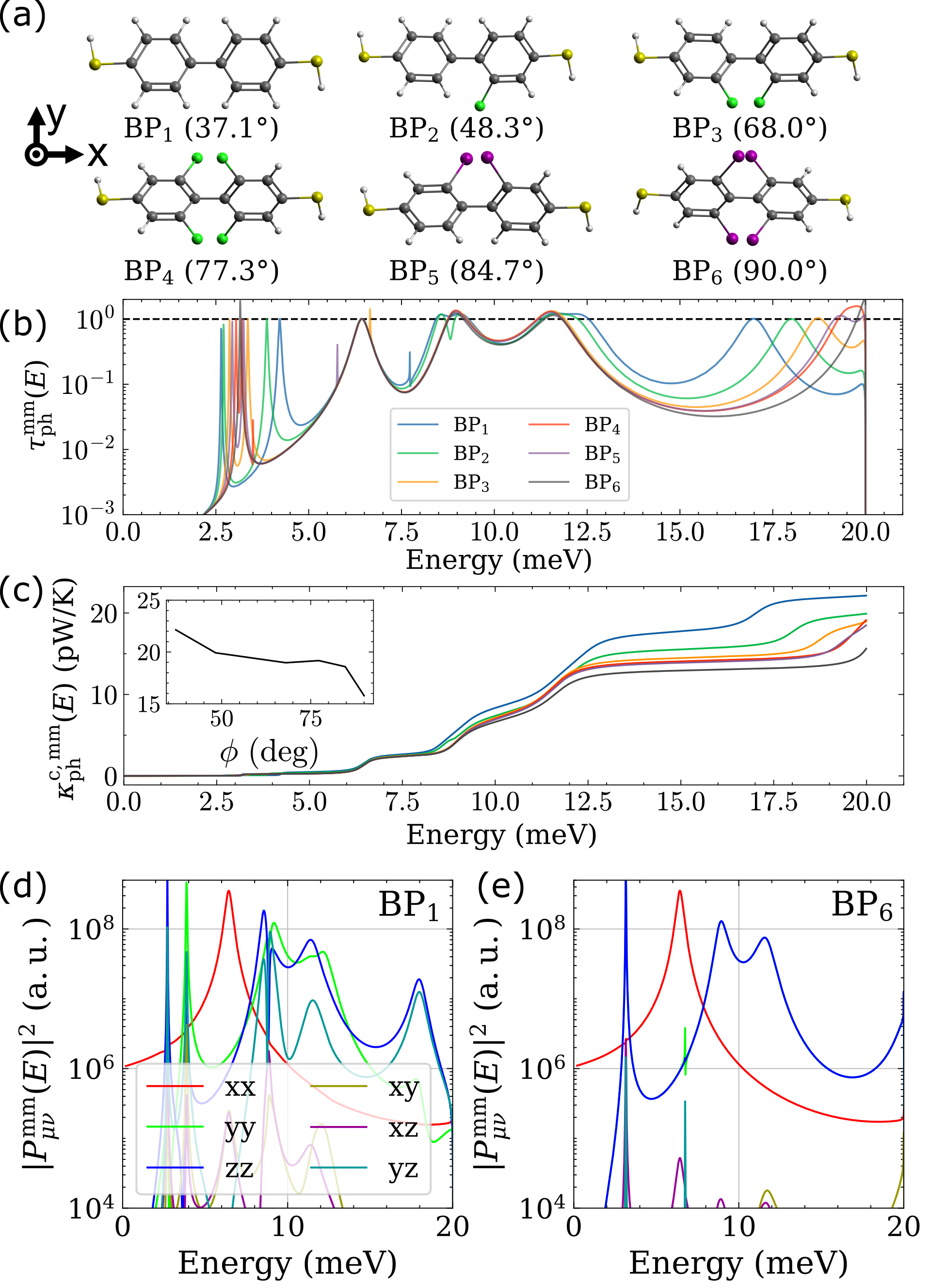}
    \caption{Transport calculations for thiol-terminated biphenyl molecules. (a) Relaxed molecular structures with corresponding dihedral angle. A coordinate system indicates the molecular orientation. (b) Phonon transmission as a function of energy. Masses of the halogen substituents are set to the hydrogen mass. 
    (c) Mass-manipulated cumulative phonon thermal conductance as a function of energy evaluated at $300~\mathrm{K}$. Thermal conductance values at $300~\mathrm{K}$ with and without mass-manipulation are given in table~\ref{tab:kappa_BP}. The inset shows $\kappa_\mathrm{ph}^\mathrm{mm}(300~\mathrm{K})$ as a function of the twist angle $\phi$.  Propagator elements from the left to right sulfur atom for (d) $\mathrm{BP_1}$ and (e) $\mathrm{BP_6}$.} 
    \label{fig:biphenyl_ED_20_S_terminated_comparison}
\end{figure}

\begin{table}[t]
\begin{tabular}{|l||l|l|l|l|l|l|}
\hline
Name & $\mathrm{BP_{1}}$ & $\mathrm{BP_{2}}$ & $\mathrm{BP_{3}}$ & $\mathrm{BP_{4}}$ & $\mathrm{BP_{5}}$ & $\mathrm{BP_{6}}$ \\ \hline \hline
Twist Angle ($\mathrm{deg}$) & 37.1 & 48.3 & 68.0 & 77.3 & 84.7 & 90.0 \\ \hline
\begin{tabular}[c]{@{}l@{}}$\kappa^{\mathrm{mm}}_{\mathrm{ph}}~(300~\mathrm{K})/\kappa_0$\end{tabular} & 22.14 & 19.92 & 18.96 & 19.17 &18.58 & 15.78 \\ \hline
\begin{tabular}[c]{@{}l@{}}$\kappa_\text{ph}(300~\text{K})/\kappa_0$\end{tabular} & 22.14 & 19.92 & 18.96 & 19.20 & 22.00 & 10.59 \\ \hline
\end{tabular}
\caption{Phonon thermal conductance for the molecules depicted in figure~\ref{fig:biphenyl_ED_20_S_terminated_comparison}(a). The table shows the thermal conductances $\kappa^{\mathrm{mm}}_{\mathrm{ph}}(300~\mathrm{K})$ and $\kappa_{\mathrm{ph}}(300~\mathrm{K})$ with and without mass-manipulation, respectively.}
\label{tab:kappa_BP}
\end{table}

The different biphenyl configurations $\mathrm{BP_1}$ to $\mathrm{BP_6}$ with dihedral angles between the phenyl rings ranging from $37.1^\circ$ to $90.0^\circ$ are depicted in figure~\ref{fig:biphenyl_ED_20_S_terminated_comparison}(a). 
The phonon transmissions in figure~\ref{fig:biphenyl_ED_20_S_terminated_comparison}(b) change with twist angle in many energy regions. This starts at the low energies between 2.5 to 5~meV, continues between 7.5 and 10~meV, 11 to 15~meV and ends at the highest energies near the Debye energy from 16 to 20~meV. 
In contrast the peak around $6.5~\mathrm{meV}$ stands out as a common feature. 

In order to clarify the underlying mechanisms, we compare the propagator elements of the two extreme cases $\mathrm{BP_{1}}$ and $\mathrm{BP_{6}}$ in figure~\ref{fig:biphenyl_ED_20_S_terminated_comparison}(d) and \ref{fig:biphenyl_ED_20_S_terminated_comparison}(e). Considering 6.5~meV first, the propagator elements exhibit a longitudinal $xx$ component for both $\mathrm{BP_{1}}$ and $\mathrm{BP_{6}}$. Due to the molecular geometry and orientation, see the coordinate system in figure~\ref{fig:biphenyl_ED_20_S_terminated_comparison}(a), this component remains largely unaffected by the rotation angle. The same propagator characteristic is observed for BP$_2$ to BP$_5$ and explains the invariance of the transmission resonance at 6.5~meV with regard to $\phi$. In the range from 2.5 to 5~meV it can be seen that two separated transmission resonances fuse to a single peak near 3~meV when $\phi$ increases to $90^\circ$. Separate transversal vibrational modes, namely in-plane transversal and out-of-plane modes that can still be well distinguished for BP$_1$ and BP$_2$ become degenerate at perpendicular orientation. The out-of-plane modes in one ring then couple to in-plane transversal modes in the other and vice versa, leading to a single transmission resonance instead of a split one. Analogous phenomena occur at 7.5 to 10~meV and 11 to 15~meV, where two rather broad maxima stemming from $zz$ and $yy$ modes fuse to a degenerate broad one of $yy$ and $zz$ character. It should be noted that the propagator elements of $yy$ and $zz$ type are degenerate for BP$_6$ basically at all energies shown in figure~\ref{fig:biphenyl_ED_20_S_terminated_comparison}(e). 
Finally, the transmission peak of BP$_1$ at around $17~\mathrm{meV}$ of $zz$ and $yz$ type is shifted to higher energies with increasing twist angle. The thermal conductance is suppressed by this mechanism, since the peak is partially shifted beyond the cutoff at the Debye energy. 

In summary, the thermal conductance generally decreases with increasing twist angle, see figure~\ref{fig:biphenyl_ED_20_S_terminated_comparison}(c). Molecule $\mathrm{BP_{4}}$ represents an exception from the ordering, because a transmission peak enters energies below the Debye-energy cutoff compared to the other configurations. BP$_6$ with $\phi=90.0^\circ$ and $\kappa^{\mathrm{mm}}_{\mathrm{ph}}(300~\mathrm{K})=15.78~\mathrm{pW/K}$ shows a reduction of $28.73\%$ compared to BP$_1$ with $\phi=37.1^\circ$ and $\kappa^{\mathrm{mm}}_{\mathrm{ph}}(300~\mathrm{K})=22.14~\mathrm{pW/K}$. The reduction in phonon thermal transport with increasing $\phi$ results from the increased coupling of in-plane transversal to out-of-plane modes, which makes the system more inhomogeneous. 

We introduced and discussed so far the mass-manipulated data to isolate the effect from twist angles. Without this mass manipulation, the results remain valid for structures with light substituents such as $\mathrm{BP_{2}}$, $\mathrm{BP_{3}}$ and $\mathrm{BP_{4}}$, see table~\ref{tab:kappa_BP}. In contrast to the fitness function, where we considered the thermal conductance at $600~\mathrm{K}$ to achieve an approximate saturation of $\kappa_\text{ph}(T)$, we used in this subsection the experimentally more relevant temperature of $300~\mathrm{K}$. As shown in figure~\ref{fig:biphenyl_ED_20_S_terminated_comparison}(b), the ordering of the thermal conductance with twist angle at $T=300$ and 600~K is the same. 

We present further studies of the twist angle dependence of the phonon thermal conductance in the Supporting Information. All of these results at various levels of theory confirm that an increasing twist angle reduces the phonon thermal conductance.

\subsubsection*{Comparison of mechanisms}
We analyze the significance of the presented mechanisms to suppress the phonon thermal conductance in the Supporting Information at the example of molecule C from figure~\ref{fig:result_overview}.  Most important in this case are (i) the terminal building blocks, reducing the thermal conductance by $93\%$, followed by (ii) the substituent effect, which suppresses the thermal conductance by $82.8\%$, (iii) para vs.\ meta coupling, yielding $40\%$  decrease, and finally (iv) the twist angle, resulting in $20\%$ reduction. The relative values specified for each mechanism denote the difference between the reference structure C and a molecular structure, where the respective mechanism has been isolated.

\section*{\label{sec:conclustions}Conclusions}
In this work, we presented a genetic algorithm to screen chemical space for molecules with lowest or highest phononic thermal conductance. The molecules were constructed from predefined building blocks, known in the literature, that are modified by halogen substituents. 
We identified important degrees of freedom of the molecular structures and allowed these degrees of freedom to be varied simultaneously by the genetic encoding. Unlike many other applications of screening methods, we employed high-level simulations for the fitness calculation. We studied the mechanisms that lead to a low or high phonon heat conductance, and pointed out four that are crucial for a suppression: (i) Specific terminal linker blocks leading to mode filtering, (ii) mass disorder and destructive interference through substituents, (iii) longitudinal and transverse mode mixing through meta instead of para couplings and (iv) molecule-internal torsion that couples in-plane to out-of-plane vibrations. We analyzed the four mechanisms systematically at different levels of theory, ranging from xTB to DFT and nearest-neighbor tight-binding approaches. Using at the same time different junction geometries and molecular configurations, we demonstrated the robustness. The highest conducting molecules instead turned out to be homogeneous linear chains without mass disorder or internal torsion. 

Overall, we showed that within our theoretical model phononic thermal conductances of covalently bonded molecules containing between $N=2$ and $4$ building blocks can be varied between $0.07$~pW/K and $33$~pW/K, yielding a variation of nearly three orders of magnitude based on molecule-internal design alone. Our study concentrated on phonons only, and electronic effects might be used to drastically increase the thermal conductance, especially of short molecules. For this reason there may be better candidates for molecules showing a high thermal conductance than what we discussed here. 

Compared to electronic behavior that may vary from metallic to insulating, yielding changes in electrical conductance of many orders of magnitude, the variations in phonon thermal transport appear to be rather moderate. This effect hence carries over from the bulk \cite{MajumdarLimit} to molecular nanostructures. The mechanisms to suppress phonon heat transport that we pointed out nevertheless provide new ideas for molecular design, which may for instance help to increase the thermoelectric efficiency of molecular devices. The genetic algorithm discovered general design principles that were not expected at the beginning. The future development of materials through inverse design methods remains exciting, and we expect many more discoveries with accelerated trial and error cycles. As a result, the impact of the developed transferable computational methodologies is much more general.

\appendix

\section*{\label{sec:Methods} Methods}

\subsection*{Genetic encoding and the size of the chemical space}

Using the encoding shown in figure~\ref{fig:encoding}, the total number of possible combinations $N$ scales exponentially with the maximum allowed molecular length. The number of combinations is determined by the number of building blocks $n_\mathrm{Block}$, the number of allowed substituents $n_\mathrm{Subs}$, the number of available couplings $n_{\text{C}}$, and the minimum and maximum number of linked building blocks $L_\text{min}$ and $L_\text{max}$, respectively. The pristine building blocks $i=1$ to $n_\mathrm{Block}=10$, shown in figure~\ref{fig:molecular_blocks}, can be modified by substituents, and each different configuration of substituents can be seen as a new building block. For each pristine block $i$ with $n_\mathrm{SubsPos}(i)$ substituent positions, we can hence derive $n_\mathrm{BlockConf}(i)=(n_\mathrm{Subs}+1)^{n_\mathrm{SubsPos}(i)}$ configurations, where the addition of 1 considers the hydrogen atom. In total, we have thus  
\begin{equation}
    n_\mathrm{B}=\sum_{i=1}^{n_\text{Block}}n_\mathrm{BlockConf}(i)
\end{equation}
derived blocks. Estimating the number of possible combinations results in:
\begin{equation}
    N \approx \sum_{i=L_\text{min}}^{L_\text{max}} n_{\mathrm{C}} \left[ n_\mathrm{B} n_\mathrm{C} \right]^i.
    \label{eq:n_combinations}
\end{equation}
We note that the approximation stems from the fact that blocks 1 and 2 do not offer $n_\text{C}$ but only one coupling position. 

Given $4$ allowed substituents (fluoride, chlorine, bromine, and iodine), for naphthalene (block 4 in figure~\ref{fig:molecular_blocks}) with $n_\mathrm{SubsPos}(i=4)=5$ we get $n_\mathrm{BlockConf}(i=4)=(4+1)^5=3125$. Using minimum and maximum lengths $L_\text{min}=2$ and $L_\text{max}=4$, two couplings $n_\text{C}=2$, i.e.\ para and meta, equation~\eqref{eq:n_combinations} yields $N\approx 8.8\times 10^{27}$ combinations. Reducing the number of substituents to $3$, still leads to $N\approx 1.3\times 10^{20}$. These huge numbers of combinations make a brute-force approach to the screening for molecules with lowest or highest phonon heat conductance impossible, but our genetic algorithm performs well even for this enormous size of chemical space.

\subsection*{Thermal conductance for fitness calculation}\label{sec:fitness_calc}

The evaluation of the phononic heat conductance for the molecular candidates provided by the genetic algorithm is challenging, as we seek the highest accuracy 
in as little time as possible. 
Several numerical schemes exist at different levels of accuracy and computational demand \cite{burkle2015first, markussen2013phonon}. 

We describe phononic transport properties as phase-coherent and elastic using Landauer-Büttiker scattering theory \cite{buttiker1988absence, Cuevas2017, wang2008quantum, wang2014nonequilibrium}. 
The phonon transmission of a molecular junction is calculated from \cite{mingo2006anharmonic}
\begin{equation}
    \tau_{\mathrm{ph}}(E) = \mathrm{Tr}\left[ \bm{G}^\mathrm{r}(E) \bm{\Gamma}_\mathrm{L}(E)\bm{G}^\mathrm{a}(E) \bm{\Gamma}_\mathrm{R}(E)\right].
    \label{eq:transmission_ph}
\end{equation}
Here, $E$ is the energy, $\bm{G}^\text{r}(E)$ ($\bm{G}^\text{a}(E)$) denotes the retarded (advanced) Green's function of the molecule and $\bm{\Gamma}_{X}(E)=-2 \mathrm{Im}\left[ \bm{\Sigma}_{X}^{\text{r}}\right]$ the linewidth broadening matrix due to the coupling to the left (L) or right (R) electrode with $X=\text{L}, \text{R}$. 
The retarded Green's function is calculated from the dynamical matrix $\bm{D}$, which is the mass-weighted Hessian of the isolated molecule, and the retarded self-energy matrices $\bm{\Sigma}^\text{r}_X(E)$ via
\begin{equation}
    \bm{G}^\text{r}(E) = \left[ (E/ \hbar)^2\bm{1}-\bm{D}-\bm{\Sigma}^\text{r}_\text{L}(E)-\bm{\Sigma}^\text{r}_\text{R}(E) \right]^{-1}.
    \label{eq:propagator}
\end{equation}

We describe the electrode within the Debye model \cite{markussen2013phonon}, which involves the following steps. The surface Green's function of the bare electrodes $g^{0,\text{r}}(E)$ is computed following Ref.~\citenum{mingo2006anharmonic}. We obtain the imaginary part from
\begin{equation}
    -\frac{1}{\pi} \mathrm{Im}\left[g^{0,\text{r}}(E) \right] = \frac{3 \hbar^2 E}{2E_\mathrm{D}^3}\Theta(E_\mathrm{D}-E).
\end{equation}
Here, $E_\text{D}$ is the Debye energy. The real part of the bare surface Green's function $g^{0,\text{r}}(E)$ is determined by a Hilbert transformation. The coupling between electrode and molecule is taken into account by solving the first-order Dyson equation with the mass-scaled force constant $\tilde{\gamma}$ to obtain the surface Green's function of the coupled electrode
\begin{equation}
    g^\text{r}(E) = g^{0,\text{r}}(E)\left[ 1 + \Tilde{\gamma} g^{0,\text{r}}(E) \right]^{-1}.
\end{equation}
Finally, the electrode embedding self-energy is determined from the surface Green's function $g^\text{r}(E)$ \cite{markussen2013phonon} via
\begin{equation}
    \left[ \bm{\Sigma}_{X}^\text{r}(E) \right]_{(i,\mu),(i,\mu)} = \Tilde{\gamma}^2 g^\text{r}(E).
    \label{eq:lead_self_energy}
\end{equation}
Only the diagonal components of the self-energy with $\mu=\{x,y,z\}$ on the atom $i$ that is directly connected to the left or right electrode are set to this value, whereas all other components of $\bm{\Sigma}_{X}^\text{r}(E)$ are assumed to vanish. Further details such as the correction of elements of $\bm{D}$ for momentum conservation can be found in the literature \cite{markussen2013phonon, mingo2006anharmonic} or in the provided code \cite{Blaschke_gaPh}. 

Based on the phonon transmission $\tau_{\mathrm{ph}}(E)$, we calculate the thermal conductance in linear response theory via
\begin{equation}
    \kappa_{\mathrm{ph}}(T)=\frac{1}{h}\int_0^{\infty} \mathrm{d}E E \tau_{\mathrm{ph}}(E)\frac{\partial n(E,T)}{\partial T}.
    \label{eq:conductance}
\end{equation}
In the expression, $T$ denotes the average temperature of left and right electrodes, and $n(E,T)=[\mathrm{exp}(E/k_\mathrm{B}T)-1]^{-1}$ is the Bose distribution function. To study the contribution of individual vibrational modes at a given temperature $T$ and energy $E$, we define the cumulative thermal conductance as
\begin{equation}
    \kappa^\mathrm{c}_{\mathrm{ph}}(E,T)=\frac{1}{h}\int_0^{E} \mathrm{d}E^\prime E^\prime \tau_{\mathrm{ph}}(E^\prime)\frac{\partial n(E^\prime,T)}{\partial T}.
\end{equation}

For electronic transport calculations, a large part of the electrode is typically added to the molecule to obtain stable transport results, forming the "extended molecule" or extended central cluster \cite{pauly2008cluster, burkle2015first}. Here, we add at most a single gold atom to each sulfur anchor, as sketched in the junction geometry in figure~\ref{fig:encoding}. The junction geometry modeled in this way closely resembles the so-called top-top geometry, widely used in similar studies \cite{Buerkle:PhysRevB2012}. Although other binding configurations are clearly possible \cite{rascon2015binding, komoto2016resolving}, we focus on molecule-internal features in this work and do not study in detail molecule-electrode interface related aspects. We calculate the thermal conductance of a molecule in a static energy-optimized geometry. Measurements on molecular junctions typically employ some form of break junction method\cite{cui2019thermal,mosso:NanoLett2019}, where mechanical stress is applied. In previous work, a decrease in conductance was observed with increasing interelectrode separation due to a reduced coupling of the molecule to the electrode\cite{cui2019thermal}. We do not explore such strain effects here, as modeling a larger electrode would be required, which does not align with the objective of computing the thermal conductance of a large population of molecules as efficiently as possible. 

The method used here, which combines Landauer-Büttiker scattering theory with xTB and may thus also be referred to as xTB-LB, is validated in the Supporting Information and shows good agreement with reference calculations from the literature \cite{klockner2017tuning}. Apart from heat transport aspects, we add the two gold atoms -- one on each side -- also for technical reasons. Namely, using solely thiol anchors, i.e.\ S-H termini, some molecules with heavy substituents attached to ethyl groups in proximity to the anchors fell apart during the xTB relaxation, invalidating these structures for applications in single-molecule transport. 

For the analysis presented in the discussion section~\ref{sec:discussion}, we sometimes nevertheless use only thiol anchors. In this case, we adjust the coupling parameter $\Tilde{\gamma}$ at the contacted sulfur atoms to obtain comparable heat transport results to the situation, when Au atoms are included. 
In particular, for Au$_1$-S termini we set the force constant to $\gamma=-7.0~\mathrm{eV}/\mathrm{\AA}^2$, and for H-S thiol anchors we use $\gamma=-1.2~\mathrm{eV}/\mathrm{\AA}^2$. These different values of force constants arise, since for Au$_1$-S the coupling is assumed to take place between two Au atoms, but for H-S between a gold and a sulfur atom. The mass-scaled force constant $\Tilde{\gamma}=\gamma/\sqrt{M_\mathrm{elec}M_\mathrm{anch}}$ is then obtained by scaling with the mass of an Au electrode atom $M_\mathrm{elec}=M_\text{Au}$ and the mass of the anchor atom $M_\mathrm{anch}$, to which the electrode atom is coupled, i.e.\ $M_\mathrm{anch}=M_\text{Au}$ for Au$_1$-S and $M_\mathrm{anch}=M_\text{S}$ for H-S. 

To optimize molecules for low conductance, an inverse relationship between fitness and phonon heat conductance is used, i.e.\ $f\propto \kappa_\text{ph}(T)^{-1}$, see equations~\eqref{eq:fitness_low-kappa}. In contrast optimization to high thermal conductance is achieved by choosing the fitness proportional to the thermal conductance, i.e.\ $f\propto \kappa_\text{ph}(T)$, see equation~\eqref{eq:fitness_high-kappa}. Further details are discussed in section theoretical approach\ref{sec:theoretical_approach} of the main text.

\subsection*{Evolution loop}
\label{sec:evolution_loop}
\begin{figure}
    \includegraphics[width=1.0\columnwidth]{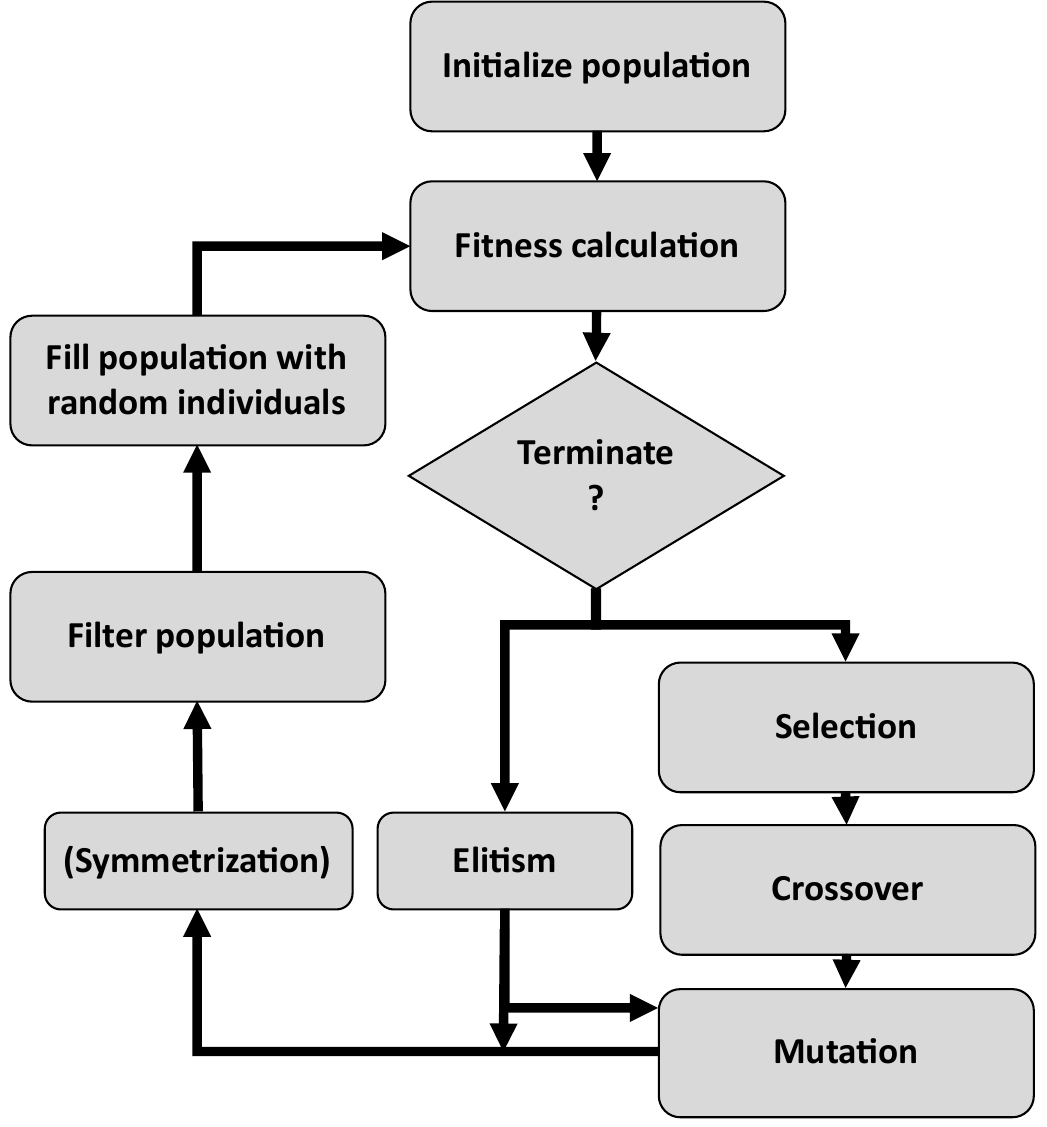}
    \caption{Flowchart of the evolution loop of the genetic algorithm.}
    \label{fig:evoloop}
\end{figure}

The evolution loop is the key ingredient of the genetic algorithm. The steps inside the loop are shown in figure~\ref{fig:evoloop} and will be discussed in detail in the following. 

The evolution is started by initializing a population with $N_\mathrm{pop}$ randomly generated individuals. Each individual represents a molecule, specified by the corresponding genetic encoding, see figure~\ref{fig:encoding}. 

After the initialization, the evolution cycle begins. First, the fitness of the individuals is computed. For this purpose, the encoding string of each individual is translated into a three-dimensional chemical structure. The structures are relaxed within the xTB framework \cite{grimme2017robust}, force constants are calculated, and the phononic heat conductance is evaluated using the techniques described in the section theoretical approach\ref{sec:theoretical_approach} and the methods part\ref{sec:fitness_calc}. The fitness is determined according to equations~\eqref{eq:fitness_low-kappa} or \eqref{eq:fitness_high-kappa} from $\kappa_\text{ph}(T)$ and other optional molecular properties such as the SP and SA. 

Subsequently, the best performing individuals are chosen to produce offspring by deterministic $k$-tournament selection \cite{tournament, miller1995genetic, 6791024}. In our scheme, the $k$ fittest individuals of each generation fill a mating pool. Offspring is produced by crossover of two parent structures A and B. The parents are uniformly sampled from the mating pool with the probability $1/k$. We implement the crossover as a length-preserving single-point crossover: A random cut is chosen in the shorter parent structure A, producing a head and a tail section. The same cut is used in parent B. The offspring is produced by combining the head of parent A with the tail of parent B and vice versa \cite{genetic_algorithm}. 

The offspring is subjected to a mutation step. A uniform random selection is made out of five different mutation operations. However, mutations are performed with a certain probability $p_\mathrm{m}$ so that an individual might go through this step without any changes. The first mutation operation is a block mutation, which replaces a building block in the encoding by a new randomly chosen block. The substituents are mapped to the new block by transferring the corresponding string sequence. If the new block has fewer substituent positions, excess ones are truncated. If there are more substituent positions, they are randomly filled in. The second mutation is a coupling mutation. The coupling at a specific position in the encoding is replaced by another random coupling. Third is the insert mutation: At a specific position, a random block with substituents and a coupling are inserted, if the length limit is not exceeded. Fourth, the truncate mutation works opposite: A block and the corresponding coupling are removed from the encoding, if the length does not fall below the previously specified lower limit. The fifth and last mutation is the substituent mutation. One substituent in the encoding is replaced by another randomly selected atom from the allowed set. Weights of the substituent types can be varied, and we set the probability for the selection of hydrogen significantly higher than for halogens. 

In addition to the candidates produced by offspring, we apply the mechanism of elitism. The $n_{\mathrm{elite}}$ best candidates are transferred directly to the next generation \cite{affenzeller2009genetic}. Additionally, those individuals are mutated with the probability $p_m$, and the resulting structures are added to the new generation without any crossover scheme, which is a non-standard operation for genetic algorithms.

In summary the next generation is formed by $n_{\mathrm{elite}}$ individuals, which are transferred directly to the next generation, and $n_{\mathrm{elite}}$ individuals, which are transferred after an additional mutation step. The remaining $N_\mathrm{pop}-2n_{\mathrm{elite}}$ individuals are generated by the described crossover and mutation operations.

Optionally, a symmetrization can be applied after the above steps. In this case the blocks of individuals are symmetrized with respect to the central block, if an odd number of blocks is present, or with respect to a central coupling, if an even number of blocks is present. The building blocks on either the left or right side of the center are randomly selected to be dominant and are correspondingly mirrored. The couplings remain unchanged, and the substituents of the old block are mapped to those of the mirrored block. The mapping works in the same way as for the block mutation. Note that since we symmetrize only the pristine building blocks but not couplings and substituents, the "symmetrized" molecules may still contain a certain degree of asymmetry. 

In the final step the population is filtered such that only unique individuals occur in the new population. Eliminated candidates are replaced by randomly generated individuals. 
The new individuals keep the population size constant and increase the genetic diversity inside the population. Our population size $N_\text{pop}$ is rather limited due to the computationally demanding fitness calculation. 

The cycle of the evolution loop is repeated until a defined number of generations has been processed. We consider the best performing molecule in the last generation as the optimal result.

\acknowledgements
We thank our experimental colleagues Marcel Mayor, Herre van der Zant, Nicolas Agra\"it, Pramod Reddy, Edgar Meyhofer and their groups for many stimulating discussions in regular meetings. We gratefully acknowledge funding by the German Research Foundation (Deutsche Forschungsgemeinschaft) within the Collaborative Research Center (Sonderforschungsbereich) 1585 (project number 492723217), subproject C02 and acknowledge use of the LiCCA high-performance computing cluster of the University of Augsburg, co-funded by the German Research Foundation (project number 499211671).

\section*{Supporting Information Available}
Validation of the xTB-LB description of phonon transport through DFT-based calculations, discussion of the length dependence of the thermal conductance for linear acetylene chains, additional studies on terminal building blocks and twist angles. Comparison of the size of all identified mechanisms to suppress heat transport at the example of molecule C from figure~\ref{fig:result_overview}(c).


\bibliographystyle{achemso_mod.bst} 
\bibliography{bibliography}

\ifarXiv


\section*{Supplementary material}

# Supporting Information for "Revealing mechanisms that control phonon heat transport through single-molecule junctions by a genetic algorithm"

Matthias Blaschke[1] and Fabian Pauly[1]

[1]*Institute of Physics and Center for Advanced Analytics and Predictive Sciences, University of Augsburg, 86135 Augsburg, Germany*
(Dated: May 25, 2025)

## S1. VALIDATION OF APPROXIMATIONS IN THE TRANSPORT APPROACH

Figure S1 compares different phonon transport methods. We study the xTB-LB scheme, used in the genetic algorithm, and the more sophisticated approach of Ref. 1 based on density functional theory (DFT), called DFT-LB.

We consider a benzene diamine molecule, which is contacted by gold electrodes. The inset of figure S1(a) shows the molecular structure with a single gold atom on both sides, as we use it in the approximate xTB-LB calculations with the genetic algorithm. The DFT studies employ instead an extended central cluster with a larger number of 20 atoms on each side. Using bromine and iodine substituents instead of hydrogen for the pink atom, we can induce destructive quantum interferences to suppress phonon heat transport.[1] Phonon transmissions are plotted in figure S1(a). Destructive quantum interferences are indeed reproduced by the xTB-LB scheme. For bromine and iodine, a dip occurs in the transmission between 16 and 20 meV, and the position of the minimum is located at a lower energy for the heavier iodine than for bromine. For hydrogen, such a destructive interference feature is absent. Our simplified xTB-LB method thus reproduces the energetic ordering of the destructive interference dips, seen in the full DFT calculations. Studying the phonon transmission further, we find that the DFT data does not show a hard cutoff at 20 meV, as imposed by the Debye energy in our xTB-LB scheme. Larger deviations are visible between DFT and xTB approaches below 10 meV. They likely arise from the different description of the electrode-molecule interface, i.e. the varied number of gold atoms in the extended central cluster.[2] However, sticking to one approach the transmission curves are similar in this low-energy region upon exchange of the pink atom of figure S1(a).

Phonon thermal conductances as a function of temperature are shown in figure S1(b). Unfortunately, the conductance of the unsubstituted diamine junction is predicted to be significantly higher by the DFT method[1] than by the xTB-LB approach applied in the genetic algorithm. For the other cases with substituents attached, the computational approaches yield good agreement. Importantly, the relative order of saturation values of the thermal conductance is the same for both methods, i.e. from high to low $\kappa_{ph}(T)$: hydrogen, bromine, iodine. This is crucial, since the genetic algorithm compares the fitness of molecules in a large population to determine optimal candidates. Absolute values are thus unimportant, but the relative grading matters.

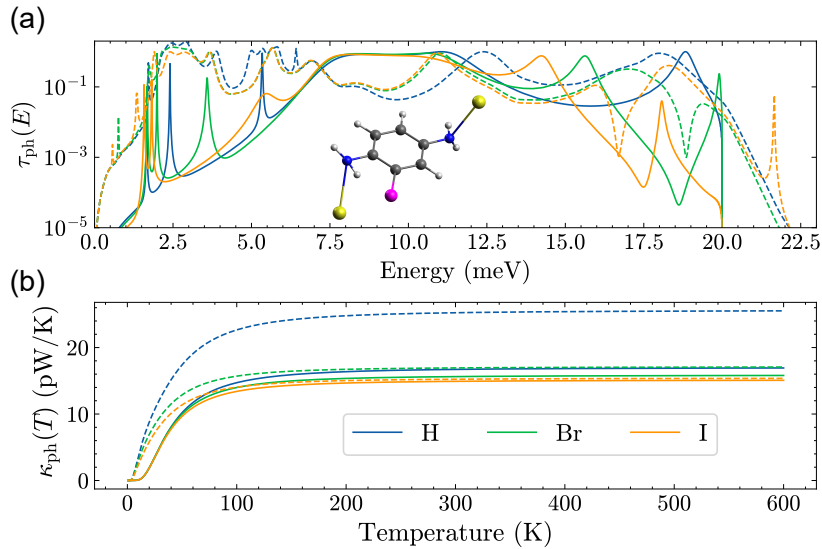

FIG. S1: (a) Phonon transmission as a function of energy for a benzene diamine molecule. The molecule is connected to gold electrodes. The inset shows the molecular structure with single gold atoms attached to the amine groups at both ends, and the pink atom is either hydrogen, bromine or iodine. Solid lines represent the results for the calculation scheme described in section A 2. Dashed lines show results, where DFT calculations and a sophisticated electrode self-energy were used.[1] (b) Corresponding phonon thermal conductance as a function of temperature.

We conclude that important molecular features such as destructive quantum interferences in the phonon transmission are reproduced by the xTB-LB scheme. Despite larger deviations in $\tau_{\mathrm{ph}}(E)$, the relative ordering of the thermal conductances $\kappa_{\mathrm{ph}}(T)$ is preserved in the saturated region for temperatures $T \gtrsim 200$ K, which is important for application in the genetic algorithm.

## S2. PHONON HEAT TRANSPORT IN LINEAR CHAINS OF ACETYLENE BUILDING BLOCKS

Optimizing for high thermal conductance in figure 4, the genetic algorithm yielded linear chains consisting of three acetylene blocks as optimal molecular structures. In these structures the gold atoms of the Au$_1$-S termini are located on the same axis as the carbon and sulfur atoms, forming a perfectly linear wire. DFT calculations show however that Au$_1$-S-C bonds are typically bent.[3] For this reason, we explore here additionally the phononic heat transport properties of chains with bent Au$_1$-S-C termini, see figure S2(a), and chains with thiol termini, see figure S2(d). Since the sulfur atoms are positioned on the same axis as the carbon atoms of the chains and since we couple the sulfur atoms directly to gold in our xTB-LB transport scheme, the molecules with thiol anchors in figure S2(d) effectively correspond to linear structures, despite the terminal hydrogen atoms being displaced from the sulfur-carbon axis. In terms of their linearity the thiolated molecules thus resemble the Au$_1$-S terminated acetylene chains studied in the main text.

By displacing the gold atoms from the axis of the carbon chain, stable molecular geometries with bent terminal Au$_1$-S-C bonds are obtained, as depicted in figure S2(a). We find these to be energetically preferred compared to the high-symmetry linear structures discussed in the main text. Figure S2(b) shows the calculated phonon transmissions as a function of energy for different chain lengths $L$, and figure S2(c) the corresponding thermal conductances. As for the linear structures reported in the main text, the chain with three acetylene blocks possesses the largest saturation value of $\kappa_{\mathrm{ph}}(T)$. The order of highest to lowest thermal conductance is $L = 3$ followed by 2, 5, 4 and 1.

Results for the related, effectively linear structures with thiol anchors are displayed in figure S2(d) to S2(f). The broadening of transmission resonances in figure S2(e) is consistently higher than for the Au$_1$-S anchors, shown in figure S2(b). Between 9 and 14 meV longitudinal and transversal modes of $xx$, $yy$ and $zz$ character together lead to transmission values exceeding 1 for the thiol-terminated molecules. Consequently, the thermal conductance of a chain with $L$ acetylenes is larger for thiol anchors, see figure S2(f), than for Au$_1$-S termini, see figure S2(c). We attribute these differences to the linear versus bent geometries of the anchors. Except for the chain with just one block, the relative order of thermal conductances from high to low with $L = 3, 1, 2, 5, 4$ for H-S is the same as for Au$_1$-S anchors. In particular, the chain with three building blocks exhibits the highest thermal conductance, which is also in agreement with the calculations presented in the main text.

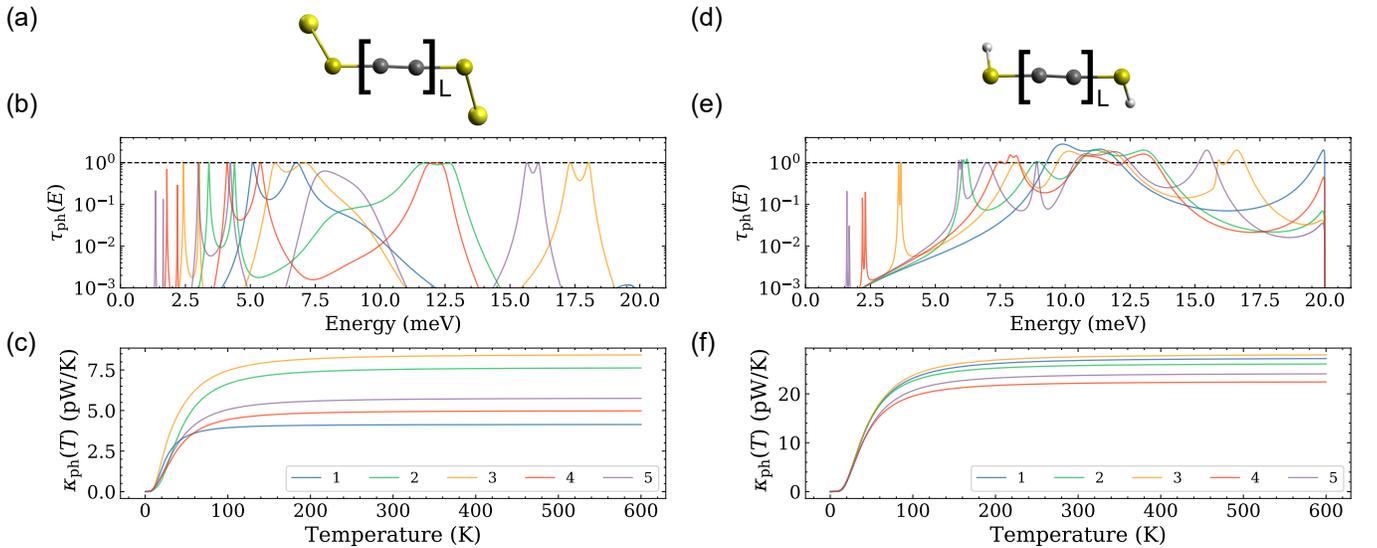

FIG. S2: Phononic transport properties of chains consisting of acetylene building blocks with Au$_1$-S and thiol anchors, respectively. (a) Structure of chains made up of $L$ acetylenes and Au$_1$-S anchors at each end. Terminal gold atoms are displaced from the chain axis. (b) Phonon transmission as a function of energy and (c) phonon thermal conductance as a function of temperature for the chains with Au$_1$-S anchors, shown in panel (a). The legend indicates the number of acetylene blocks $L$, and the color coding is identical in panels (b) and (c). (d,e,f) Same as panels (a,b,c), respectively, but for thiol anchors.

## S3. MECHANISMS TO SUPPRESS PHONONIC HEAT TRANSPORT

As presented in the main text, we have identified four mechanisms to suppress phonon heat transport with the help of the genetic algorithm. They are (i) acetylene terminal blocks, (ii) substituents, (iii) meta coupling, (iv) molecule-internal torsion. We will in the following discuss in more detail the mechanisms (i) and (iv). Different computational approaches will be used to demonstrate their robustness. Finally we will compare the size of all four mechanisms in order to assess their relative importance.

### A. Additional model for silicon electrodes

In the methods part "Thermal conductance for fitness calculation" of the main text, we have discussed our model to couple $Au_1$-S and H-S termini to gold electrodes. In order to explore also higher energy regions, we will consider silicon electrodes in the tight binding models discussed below. Following Ref. 4, we use a Debye energy of $E_D = 70$ meV and couple directly from the silicon electrode to a carbon atom of the molecule, setting the force constant to $\gamma = -4.0$ eV/Å$^2$. The masses for the mass scaling are thus $M_{elec} = M_{Si}$ for the electrode and $M_{anch} = M_C$ for the anchor. Due to the significantly higher Debye energy, we analyze the thermal conductance for Si electrodes typically at a temperature of $T = 800$ K to reach a saturated value of $\kappa_{ph}(T)$.

### B. Mechanism 1: Terminal building blocks

To achieve a better understanding, we study the influence of terminal building blocks on phonon heat transport using a nearest-neighbor carbon model, thus disregarding hydrogen atoms. The model is depicted in figure S3(a) for a benzene molecule with attached acetylene terminal building blocks, called $B_{ac}$. We describe the nearest-neighbor interaction between carbon atoms by extracting the principal components of the carbon-carbon coupling from GFN1-xTB calculations. As shown by the blue, dark-grey and red colors of the bonds in figure S3(a), we distinguish single, double and triple carbon-carbon bonds. We extract the principal components of these bonds from calculations on ethane, phenyl and acetylene, respectively. All of the blue single-bond couplings are characterized by the coupling matrices $d_{ij}$ between neighboring atoms $i$, $j$, containing the mass-scaled force constants $\tilde{\delta}_{xx}$, $\tilde{\delta}_{yy}$ and $\tilde{\delta}_{zz}$ as diagonal components. Similarly, for the dark-grey double bonds inside the phenyl ring, the coupling matrix is called $d'_{ij}$, formed by $\tilde{\delta}'_{xx}$, $\tilde{\delta}'_{yy}$ and $\tilde{\delta}'_{zz}$ as diagonal components, and for the triple bonds they are $d''_{ij}$ with diagonal entries $\tilde{\delta}''_{xx}$, $\tilde{\delta}''_{yy}$ and $\tilde{\delta}''_{zz}$, see figure S3(a). Since bonds may be aligned differently in the $x$-$y$ plane, the nonvanishing coupling components $D_{ij} = R_z^T(\theta_{ij}) Y_{ij} R_z(\theta_{ij})$ between neighboring atoms $i$, $j$ of the full dynamical matrix $D$ are finally obtained by applying appropriate transformations to $Y_{ij} = d_{ij}, d'_{ij}, d''_{ij}$. Here, $R_z(\theta_{ij})$ describes the rotation around the $z$-axis by the angle $\theta_{ij}$ for each coupling or the identity matrix, if the angle vanishes, see figure S3(a). In a last step, the acoustic sum-rule is enforced by setting the diagonal elements of the dynamical matrix to $D_{\alpha\alpha} = -\sum_{\alpha \neq \beta} D_{\alpha\beta}$, where the indices $\alpha, \beta$ describe both atomic and Cartesian components.

We hypothesize that the difference in force constants between the dark-grey bonds of the phenyl ring and the red and blue bonds of the acetylene terminal building blocks in figure S3(a) may be the reason for the suppressed phonon heat transport for molecules with attached acetylene linkers. To check this assumption, we compare the transport through $B_{ac}$ using the mixture of mass-scaled force constants $\tilde{\delta}_{\mu\mu}, \tilde{\delta}'_{\mu\mu}, \tilde{\delta}''_{\mu\mu}$ with $\mu = x, y, z$, as described above, to the hypothetical case where the force constants in the acetylene bonds match those inside the phenyl ring. In this hypothetical case, we use only the force constants $\tilde{\delta}'_{\mu\mu}$, i.e. $d_{ij}$ and $d''_{ij}$ are replaced by $d'_{ij}$. We refer to this homogeneous model as $hB_{ac}$.

Employing silicon electrodes, the transmission is shown in figure S3(b) and the cumulative thermal conductance in figure S3(c). Transport is indeed strongly enhanced, when the same force constants are chosen for terminal blocks and the phenyl ring. In the energy range from 0 to 50 meV, the transmissions of $B_{ac}$ and $hB_{ac}$ in figure S3(b) are rather similar. Transmission resonances appear to be somewhat shifted, and the transmission of $B_{ac}$ is typically below those of $hB_{ac}$ except for smaller energy intervals where $hB_{ac}$ features transmission resonances. In the region between 50 and 57 meV the transmission of $B_{ac}$ is strongly suppressed compared to $hB_{ac}$. This is also the case for higher energies except for the places where the transmission of $B_{ac}$ shows rather narrow resonances.

On the scale in figure S3(c), the cumulative thermal conductances of $B_{ac}$ and $hB_{ac}$ hardly differ in the energy range from 0 to 26 meV. Between 26 and 50 meV the cumulative thermal conductance of $hB_{ac}$ grows slightly faster than that of $B_{ac}$ until it increases much faster above 50 meV. Studying the thermal conductance at the saturation value of $T = 800$ K, it is enhanced by approximately 90% for $hB_{ac}$.

From the energy-dependent plots of the propagator in figure S3(d) we see that the lowest transmission resonances near 10 meV originate from out-of-plane $zz$ modes for both models, the subsequent ones near 14 meV from longitudinal $xx$ modes. In-plane transversal modes of $yy$ type are suppressed at 14 meV for $B_{ac}$ as compared to its homogeneous counterpart. In the energy range between 16 and 50 meV the most important difference is the suppression of in-plane transversal modes of $yy$ type for $B_{ac}$, and for certain energy intervals also of vibrations of $zz$ type. In the interval of 50 to 57 meV we find contributions of all kinds of

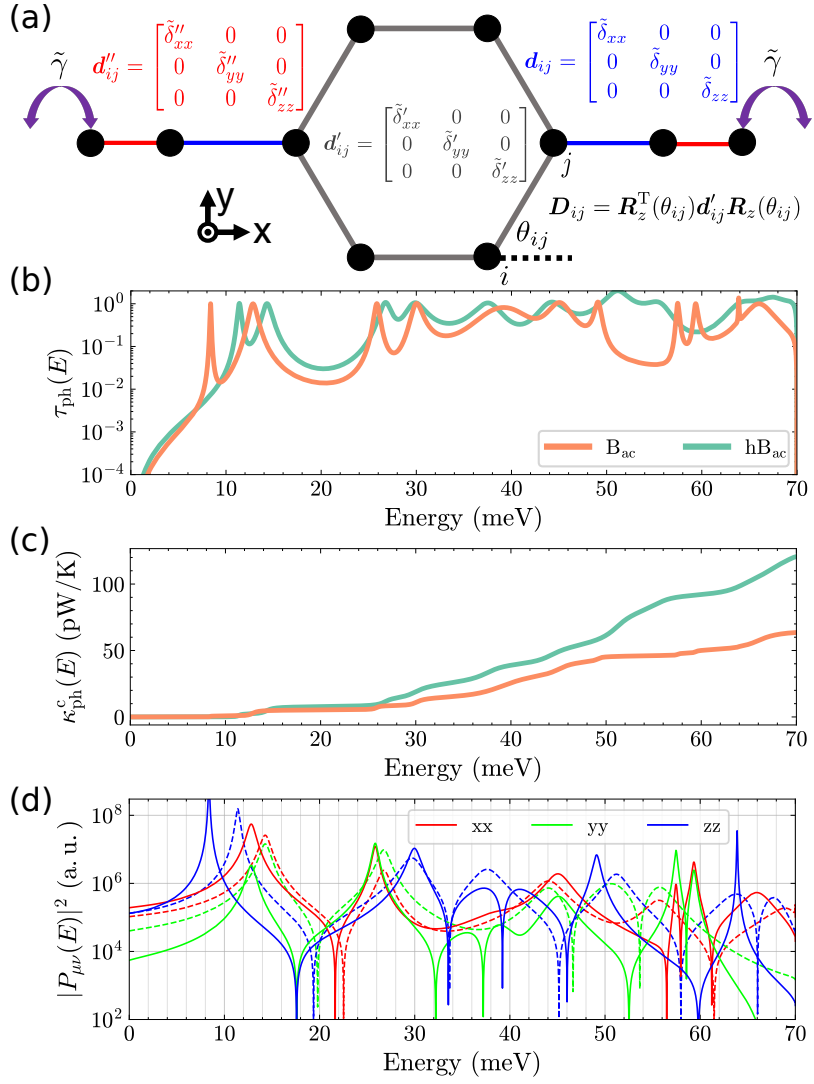

FIG. S3: (a) Nearest-neighbor coupling model for $B_{ac}$, i.e. a benzene molecule with acetylene terminal building blocks that are aligned in para configuration. The dynamical matrix is constructed from the nearest-neighbor coupling matrices $\boldsymbol{d}_{ij}$ for the blue single bonds, $\boldsymbol{d}'_{ij}$ for the dark-grey double bonds and $\boldsymbol{d}''_{ij}$ for the red triple bonds with $i, j$ being nearest neighbor atoms. Appropriate transformations with $\boldsymbol{R}_z(\theta_{ij})$ consider rotations of bonds in the $x$-$y$ plane, as shown exemplarily for $\boldsymbol{d}'_{ij}$. The rotations $\boldsymbol{R}_z(\theta_{ij})$ around the $z$ axis are performed by the corresponding angle $\theta_{ij}$, measured with respect to the $x$-axis. The force constants, extracted from GFN1-xTB, yield the mass-scaled principal components $\tilde{\delta}_{xx} = -1.74$ eV/($\text{Å}^2$u), $\tilde{\delta}_{yy} = -0.58$ eV/($\text{Å}^2$u), $\tilde{\delta}_{zz} = -0.58$ eV/($\text{Å}^2$u), $\tilde{\delta}'_{xx} = -2.62$ eV/($\text{Å}^2$u), $\tilde{\delta}'_{yy} = -0.87$ eV/($\text{Å}^2$u), $\tilde{\delta}'_{zz} = -0.39$ eV/($\text{Å}^2$u), and $\tilde{\delta}''_{xx} = -9.62$ eV/($\text{Å}^2$u), $\tilde{\delta}''_{yy} = -0.10$ eV/($\text{Å}^2$u), $\tilde{\delta}''_{zz} = -0.10$ eV/($\text{Å}^2$u), determining the diagonal matrices $\boldsymbol{d}_{ij}, \boldsymbol{d}'_{ij}, \boldsymbol{d}''_{ij}$ respectively. The realistic configuration $B_{ac}$ is compared to the homogeneous model $hB_{ac}$, where $\tilde{\delta}_{\mu\mu}$ and $\tilde{\delta}''_{\mu\mu}$ are set to the values of $\tilde{\delta}'_{\mu\mu}$ with $\mu = x, y, z$. (b) Phonon transmission as a function of energy for silicon electrodes. The orange curve shows the behavior of $B_{ac}$ and the green curve those of $hB_{ac}$. (c) Cumulative thermal conductance as a function of energy evaluated at 800 K. (d) Terminal propagator elements as a function of energy for $B_{ac}$, shown with solid lines, and $hB_{ac}$, shown with dashed lines.

transversal and longitudinal modes, namely of type $xx$, $yy$ and $zz$, to be lower for $B_{ac}$. For energies above 62 meV, mainly $zz$ and $yy$ transversal vibrations are attenuated for $B_{ac}$ as compared to $hB_{ac}$.

In order to rule out systematic errors in our xTB-LB transport simulations, we have computed heat transport through benzene, B, and benzene with acetylene units, $B_{ac}$, using DFT. The molecules are attached to gold electrodes through sulfur atoms, as shown in figure S4. Phonon transport is calculated as described in Ref. 6. Atomically sharp gold tips of 20 atoms are included in the extended central cluster, attached to both sides of the molecules. Technically, we obtain the Hessian matrix for the molecular junctions using TURBOMOLE,[7] employing the PBE exchange-correlation functional[8] and the def-SV(P) Gaussian basis set[9] for all atoms. We converged the gradient norm to below $10^{-5}$ a.u. during geometry optimization and the energies to below $10^{-8}$ a.u..

The energy-dependent transmission is depicted in figure S4(a) and the cumulative thermal conductance at 300 K in figure S4(b). Similar to the GFN1-xTB-based calculations in the main text, see figure 6, and the model calculations in figure S3,

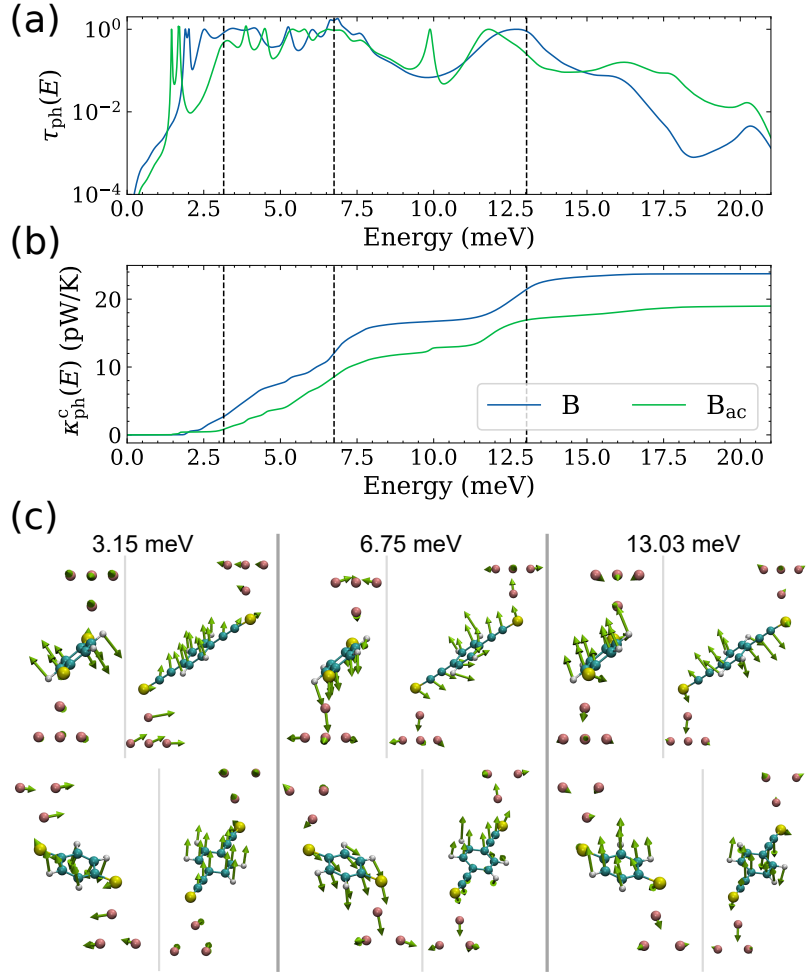

FIG. S4: Phonon transport through benzene, B, and benzene with acetylene terminal building blocks, $B_{ac}$, using a DFT-based approach. The molecules are connected to gold electrodes through sulfur atoms. (a) Phonon transmission as a function of energy. (b) Cumulative thermal conductance as a function of energy, evaluated at $T = 300$ K. The thermal conductance at room temperature is $\kappa_{ph}(300$ K$) = 23.78$ pW/K for B and $\kappa_{ph}(300$ K$) = 19.01$ pW/K for $B_{ac}$. (c) Dominant transmission eigenchannel for the molecular junction formed by B in the left column and $B_{ac}$ in the right column, respectively, at energies of 3.15 meV, 6.75 meV and 13.03 meV.[5] Pictures at the top and bottom show the specific transmission eigenchannel from different perspectives. Phonon waves enter from the upper electrode. The energies selected for the plots of the eigenchannels are indicated by vertical dashed lines in panels (a) and (b).

we find that phononic heat transport is reduced by the acetylene terminal building blocks. The suppression in the DFT model is however somewhat less than in the studies with GFN1-xTB and the nearest-neighbor tight-binding model. The cumulative thermal conductance indicates that decisive phonon energies are located at around 3, 7 and 13 meV, where $\kappa_{ph}^c(E)$ increases more strongly for B than for $B_{ac}$.

Phonon transmission eigenchannels are plotted in figure S4(c) at those energies that are marked in figure S4(a) and S4(b) by vertical dashed lines. For B we find a combination of in-plane transversal and out-of-plane transversal character at 3.15 meV, which is changed to mainly longitudinal and out-of-plane character for $B_{ac}$. At 6.75 meV the eigenchannels of B and $B_{ac}$ both show mixed longitudinal and out-of-plane character. At 13.03 meV, finally, the molecular junction containing B exhibits predominantly longitudinal and out-of-plane characteristics, which is modified to basically out-of-plane type for $B_{ac}$. To summarize, we note that the complex contact geometries in the DFT simulations complicate the analysis why phonon transport is suppressed for $B_{ac}$ as compared to B. This justifies the study of the simplified nearest-neighbor models discussed before to obtain a basic understanding.

To corroborate our findings for the ethyl groups, we present a comparison of the Cartesian propagator elements for all anthracene-based structures, A, $A_{ac}$ and $A_{et}$, shown in figure 6. The results are depicted in figure S5. In agreement with our discussion in the main text, the $xx$-components are similar for all three structures. The peaks around 6 meV exhibit comparable widths. For $A_{et}$ we find an additional feature around 4 meV, which is however insignificant for the thermal conductance due to its small width. In the $yy$-components, a strong suppression of modes with energies larger than 7 meV can be observed for both $A_{ac}$ and $A_{et}$. This suppression is more pronounced for $A_{et}$ than for $A_{ac}$. For the $zz$-components, $A_{et}$ shows a suppression of

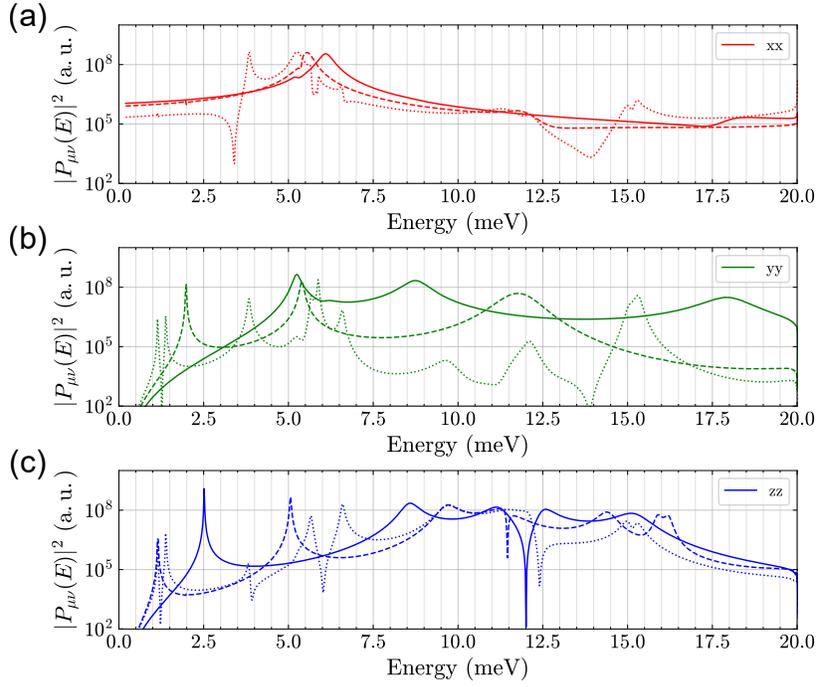

FIG. S5: Absolute square of the propagator elements $|P_{\mu\nu}(E)|^2$ for the anthracene-based molecules from figure 6, resolved into (a) $xx$, (b) $yy$ and (c) $zz$ Cartesian components. A: solid lines; A$_{ac}$: dashed lines; A$_{et}$: dotted lines.

modes at 2.5 meV and between around 13 and 16 meV compared to A. In summary, the ethyl end group induces a suppression of both $yy$- and $zz$-modes, particularly at high energies.

## C. Mechanism 4: Twist angle

Similar to subsection S3 B on terminal building blocks, we analyze the dependence of heat transport on the twist angle using a nearest-neighbor tight-binding model. We choose biphenyl as the prototypical molecule[3,10–12] and distinguish single from double carbon-carbon bonds, as shown in figure S6(a). The matrices $d_{ij}$ and $d'_{ij}$, describing the coupling between neighboring carbon atoms $i, j$ for single and double bonds, respectively, are the same as in the previous paragraph. The mixing between in-plane transversal and out-of-plane modes of the planar system is identified as the important mechanism that suppresses the phonon thermal conductance due to a molecule-internal twist. To focus on this effect, we set the principal components $\tilde{\delta}_{xx}$ and $\tilde{\delta}'_{xx}$ to zero, eliminating longitudinal vibrations with respect to the bond direction. In addition to the transformations of $d'_{ij}$ by rotation matrices $R_z(\theta_{ij})$ to account for the different alignment of double bonds of the phenyl rings in the $x$-$y$ plane, we need to consider now the twist angle $\phi$. For this purpose, the coupling matrices $d'_{ij}$ are subsequently rotated around the $x$-axis by $R_x(\phi)$, if both atoms $i$ and $j$ are located in the ring on the right side of figure S6(a). Since $\tilde{\delta}_{yy} = \tilde{\delta}_{zz}$, no $x$-axis rotation is needed for the single bonds, described by $d_{ij}$. The other procedures to compute phonon heat transport within the tight-binding model are identical to section S3 B.

The transmission $\tau_{\mathrm{ph}}(E)$ is shown as a function of energy in figure S6(b). For $\phi = 0°$ it exhibits split peaks in several energy regions, namely between 12 to 16 meV, 20 to 26 meV, 44 to 48 meV and 56 to 64 meV, which fuse to a single one with increasing $\phi$. For the features above 30 meV, a decrease in transparency is additionally apparent as $\phi$ grows.

Analysis of terminal propagator elements, exemplified for $\phi = 0°$ and $90°$ in figure S6(c), reveals the character of the phonon modes. The ring rotation mixes the in-plane transversal modes of the planar system with the out-of-plane transversal vibrations, and $yy$ and $zz$ modes of the propagator thus become perfectly degenerate at $\phi = 90°$. Since $zz$ and $yy$ modes exhibit significantly different force constants on the rings ($\tilde{\delta}'_{yy}/\tilde{\delta}'_{zz} \approx 2$), the resulting mismatch leads to reduced transparency. The split peaks between 20 to 26 meV and 44 to 48 meV are seen to arise from $yy$ and $zz$ vibrations of the planar systems. Due to the mixing of the transversal modes, they fuse to a single peak, which is energetically positioned in the middle of the original $yy$ and $zz$ vibrational energies. The double maxima in the regions of 12 to 16 meV and 56 to 64 meV are seen to be of a different origin, stemming from $yy$ and $zz$ modes, respectively. Consistent with the strongly reduced transmissions above 30 meV in figure S6(b), maxima of the propagator elements for $\phi = 90°$ are much lower than for $\phi = 0°$ at these high energies, see figure S6(c).

The cumulative thermal conductance in figure S6(d) increases rather stepwise. The steps are determined by the transmission

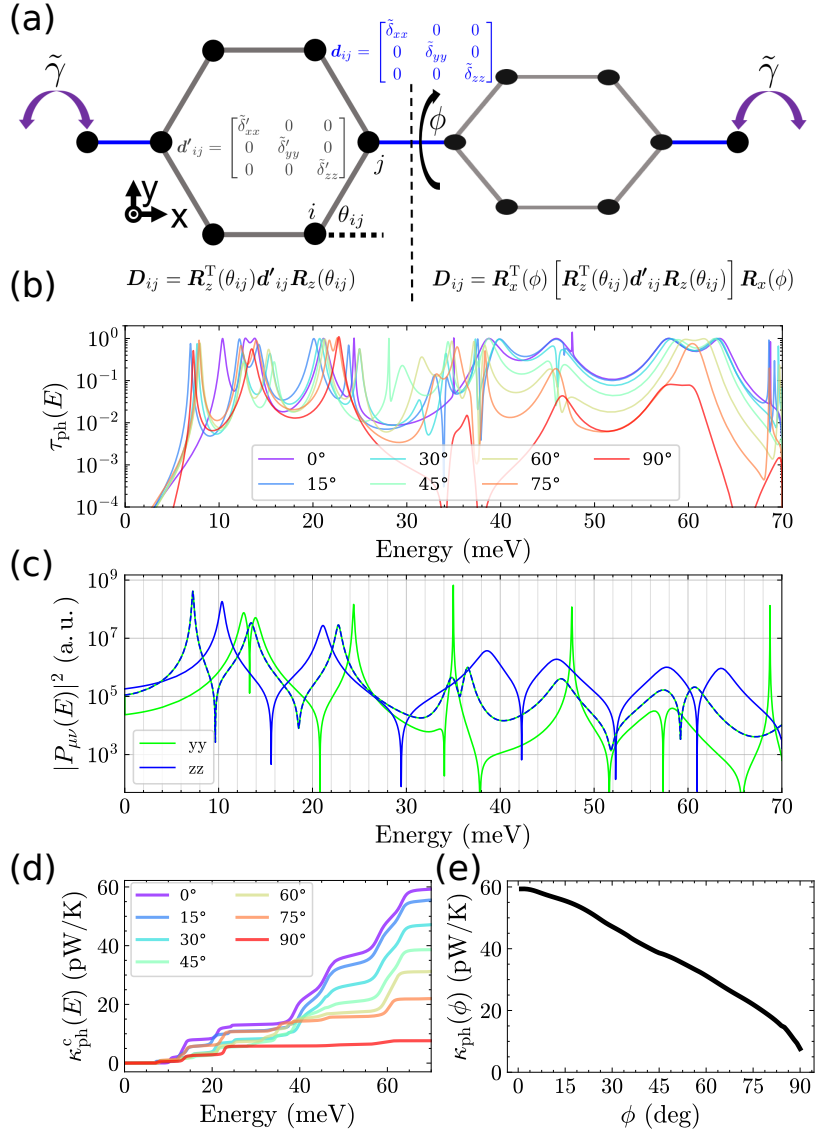

FIG. S6: (a) Nearest-neighbor coupling model for biphenyl. Distinguishing single from double bonds, the couplings between neighboring carbon atoms are described by $\boldsymbol{d}_{ij}$ and $\boldsymbol{d}'_{ij}$, respectively, with principal components $\tilde{\delta}_{\mu\mu}, \tilde{\delta}'_{\mu\mu}$ and $\mu = x, y, z$. The values of the principal components are specified in the caption of figure S3. To focus on transversal vibrations, we set $\tilde{\delta}_{xx} = \tilde{\delta}'_{xx} = 0$ here. To account for the different orientation of bonds between atoms $i$ and $j$, $\boldsymbol{d}'_{ij}$ is rotated around the $z$ axis by the corresponding angle $\theta_{ij}$, measured with respect to the $x$-axis. When both atoms $i$ and $j$ are positioned on the right ring, the matrix $\boldsymbol{d}'_{ij}$ is additionally rotated around the $x$-axis by the twist angle $\phi$. The molecule is contacted to the electrodes at the left and right sides, as shown by the purple arrows. (b) Transmission as a function of energy for the indicated twist angles $\phi$ using silicon electrodes. (c) Relevant terminal propagator elements of the diagonal Cartesian components for $\phi = 0°$, shown with solid lines, and $\phi = 90°$, shown with dashed lines. (d) Cumulative thermal conductance at 800 K plotted against energy for the indicated twist angles $\phi$. (e) Thermal conductance at 800 K as a function of the twist angle $\phi$.

resonances, which decrease in width and height, as $\phi$ increases. This leads to the reduced step heights in $\kappa_{\mathrm{ph}}^{\mathrm{c}}(E)$ at the relevant vibrational energies for increased twist.

Figure S6(e) depicts the phonon heat conductance at 800 K as a function of the twist angle $\phi$. It nicely shows how $\kappa_{\mathrm{ph}}(800\,\mathrm{K})$ decreases almost linearly with increasing $\phi$ for our model without the longitudinal modes. We conclude that the twist angle is a robust mechanism for controlling the phononic thermal conductance. The mechanism is a coupling of in-plane and out-of-plane transversal modes of the planar molecule that are indistinguishable at perpendicular ring alignment. The mismatch of transversal force constants of the carbon double bond causes the suppression.

In addition to the preceding nearest-neighbor tight-binding study and the analysis in the main text, we investigate the effect of the twist angle in DFT calculations. We compute the Hessian matrix for two biphenyl-derived molecular junctions using TURBOMOLE.[7] Technically, we employ the PBE exchange-correlation functional[8] and the def-SV(P) Gaussian basis set[9] for all atoms. The gradient norm is converged to better than $10^{-5}$ a.u. during geometry optimization, and the energies to better than

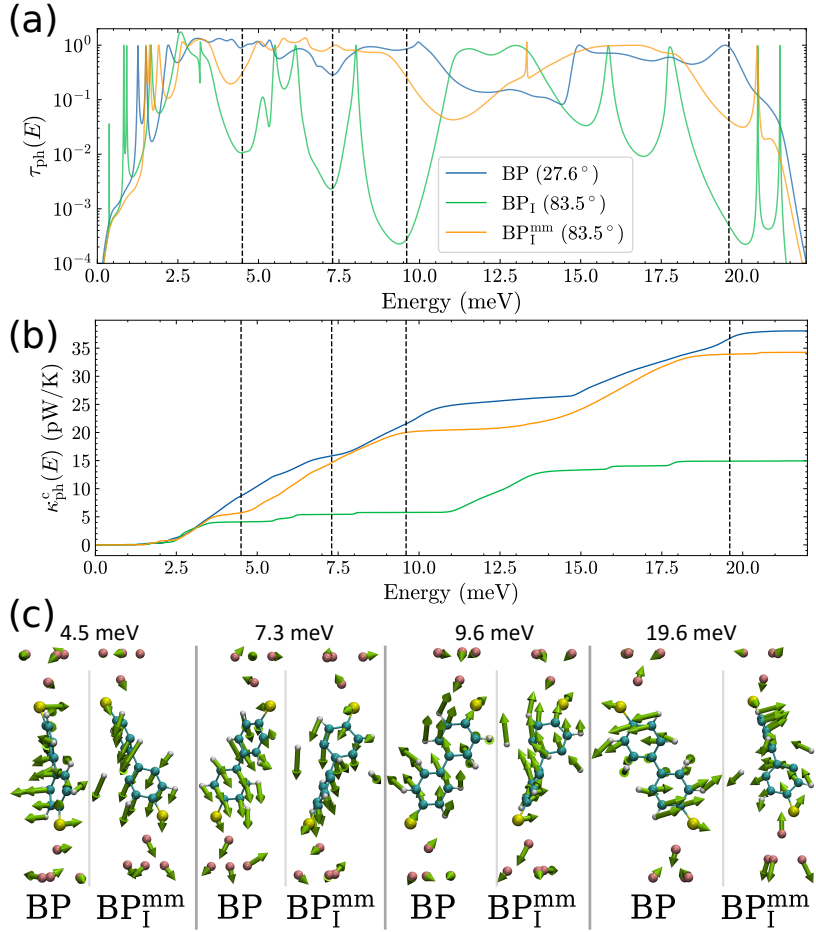

FIG. S7: (a) Transmission as a function of energy for the molecular junctions indicated in the legend. BP is the regular biphenyl, $BP_I$ the biphenyl with four iodine substituents attached, and $BP_I^{mm}$ is $BP_I$ with the mass of the iodine substituents set to the hydrogen mass. Dihedral angles of the biphenyls are indicated in the legend. (b) Cumulative thermal conductance as a function of energy at 300 K for the different biphenyl junctions. (c) Phonon transmission eigenchannels of BP and $BP_I^{mm}$ for phonon waves entering from the upper electrode. The selected energies are marked by vertical dashed lines in panels (a) and (b).

$10^{-8}$ a.u.. Phonon transport is determined as described in Ref. 6. The extended central cluster contains the biphenyl derivative, which is connected by terminal sulfur atoms to atomically sharp gold tips of 20 atoms at both sides.

Our reference structure is the molecular junction containing the unsubstituted biphenyl molecule, BP (similar to $BP_1$ in figure 8), featuring a dihedral angle of $\phi = 27.6°$. To induce a twist angle we attach four iodine substituents in the ortho positions of the ring-connecting carbon atoms of biphenyl in the second structure, called $BP_I$ (similar to $BP_6$ in figure 8). They lock the phenyl rings at a dihedral angle of $83.5°$. To isolate the influence of the twist angle, we set the mass of the iodine substituents to that of hydrogen while keeping all atoms in the positions determined for $BP_I$. We refer to this molecular junction as $BP_I^{mm}$.

Calculated transmissions and the corresponding cumulative thermal conductances for BP, $BP_I^{mm}$ and $BP_I$ are depicted in figure S7(a) and S7(b), respectively. Compared to BP, the phonon thermal conductance at 300 K is suppressed by 10% for $BP_I^{mm}$ with increased twist angle in accordance with the previous findings. Without mass manipulation, the reduction of the thermal conductance for $BP_I$ even amounts to 59%. Figure S7(b) reveals the most important energy regions, where the cumulative thermal conductance of BP exceeds those of $BP_I^{mm}$ and $BP_I$.

To understand how heat is transported through the molecular junctions by vibrations, we study phonon transmission eigenchannels.[5] The dominant first eigenchannels are presented for BP and $BP_I^{mm}$ in figure S7(c) at selected energies. Second or higher eigenchannels yield negligible contributions and are therefore not shown. At 4.5 meV, BP exhibits an out-of-plane character. In contrast, $BP_I^{mm}$ shows this behavior only in the upper ring whereas the atoms on the lower phenyl ring move in in-plane direction. This leads to a reduced transparency, see figure S7(a), due to the coupling of modes with different force constants. A similar situation is evident in figure S7(c) at 9.6 meV and 19.6 meV. In these cases, BP exhibits a transverse in-plane mode, but $BP_I^{mm}$ shows this behavior only in one ring, whereas the second ring features an out-of-plane motion. At both energies, the transmissions values of $BP_I^{mm}$ are suppressed compared to BP. For cross-validation, we analyze the eigenchannels

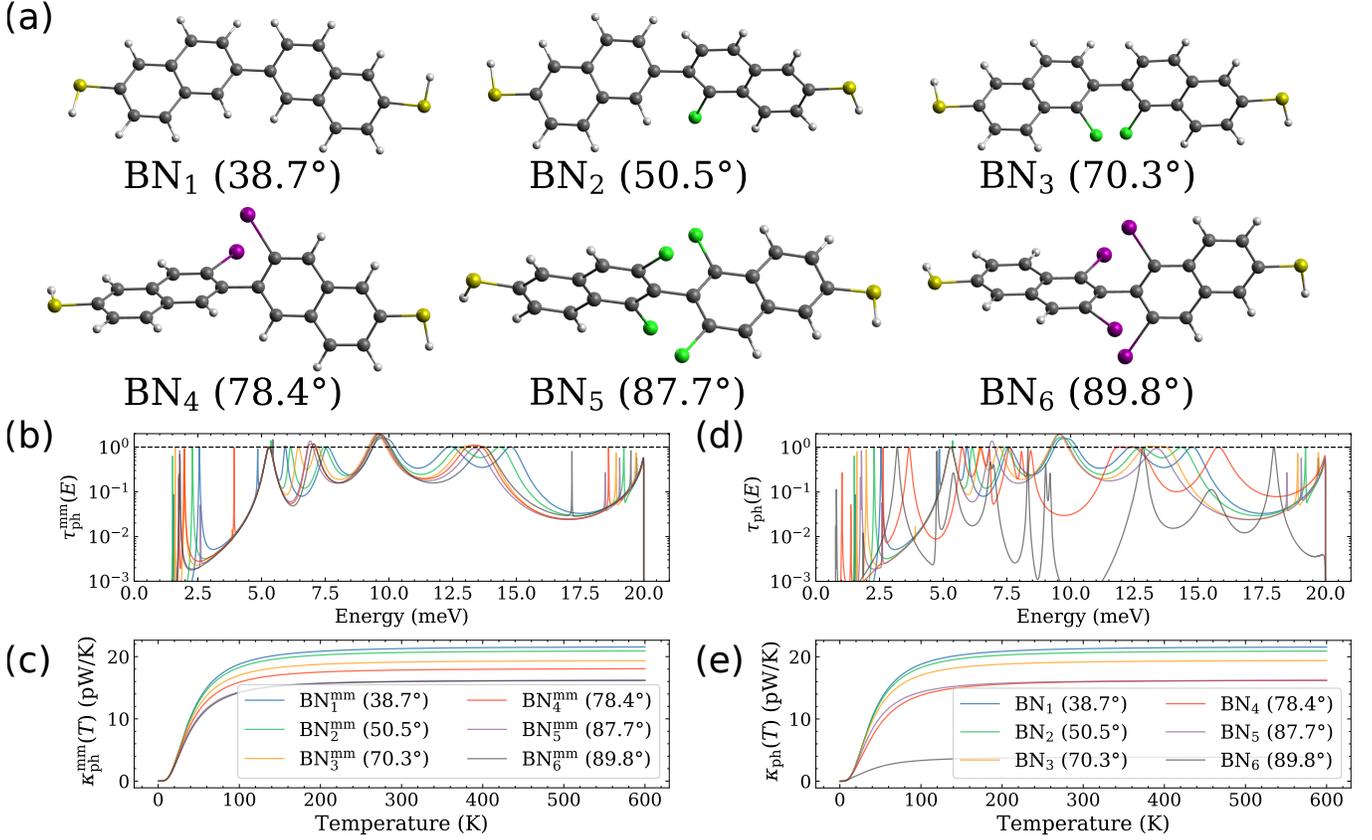

FIG. S8: Phononic heat transport calculations through thiol-terminated binaphthalene molecules with different dihedral angles, attached to gold electrodes. (a) Optimized molecular geometries. The dihedral angle between the naphthalene units is given in brackets. (b) Transmission as a function of energy for the different molecular junctions, when masses of the halogen substituents are set to the hydrogen mass. (c) Thermal conductance as a function of temperature for the various binaphthalene derivatives with mass manipulation. (d) and (e) Same as in (b) and (c), respectively, but without mass-manipulation.

at 7.3 meV. Here, $BP_1^{mm}$ exceeds the transmission of BP, which is the opposite trend compared to the three energies discussed so far. At 7.3 meV both molecules feature an in-plane longitudinal movement where the effects of ring rotation are indeed expected to be small.

We have shown that an increasing twist angle from planar to perpendicular reduces the phonon thermal conductance. For $BP_1^{mm}$ we have eliminated the influence of the side groups apart from their steric effects on $\phi$. The thermal conductance of $BP_1$ without mass manipulation is suppressed even further, see figure S7(b). As discussed in the context of table 2 in the main text, phonon thermal conductances with and without mass manipulation may be identical for light substituents. In contrast to the heavy iodine, the effect of molecule-internal twist on vibrational heat transport may thus be demonstrated for light substituents without a mass manipulation. The effect might be used to realize a controllable switch for thermal transport, if the molecular twist angle in a junction is adjusted by external stimuli like electric fields[13,14] or charge on the molecule.

To generalize the results beyond biphenyl molecular junctions, we study phonon heat transport through binaphthalenes. As shown in figure S8(a), the naphthalene units are locked in different dihedral angles by halogen substituents.

To isolate the effect of the twist angle, we manipulate again the masses of the halogen substituents for the analysis of transmissions and temperature-dependent thermal conductances in figure S8(b) and S8(c). The most significant deviations in transmission values are observed for the different molecules between 11 and 15 meV. $BN_1^{mm}$ and $BN_2^{mm}$ exhibit two peaks in this energy range which merge to a single one with increasing twist angle. Additionally the maximum transmission decreases in this energy range for growing $\phi$. Figure S8(c) shows that the thermal conductance decreases with increasing twist angle between the naphthalene units. The values of $\kappa_{ph}(T)$ for $BN_5^{mm}$ and $BN_6^{mm}$ nearly lie on top of each other as $\phi$ is almost the same. When comparing $BN_1^{mm}$ and $BN_6^{mm}$, the thermal conductance at 300 K is reduced by 24%, which is of the same order of magnitude as for the biphenyl molecules $BP_1^{mm}$ and $BP_6^{mm}$ studied in the main text.

Since large mass manipulations are not practical for real molecules, figure S8(d) and S8(e) presents transmissions and thermal conductances without mass manipulation. Surprisingly, hardly any difference can be seen for the structures $BN_1$, $BN_2$, $BN_3$, $BN_5$ in comparison to figure S8(b) and S8(c), respectively. Those molecules are locked by chlorine substituents, which are significantly lighter than the iodines used in $BN_4$ and $BN_6$. The transport characteristics of $BN_4$ and $BN_6$ in contrast exhibit

considerable changes to $BN_4^{mm}$ and $BN_6^{mm}$, respectively. For $BN_4$, several new transmission resonances emerge, and $BN_6$ features a particularly low transmission with narrow resonances across the whole energy range studied. Despite these differences, the relative order of the thermal conductances in figure S8(e) is the same as in figure S8(c). For $BN_6$, the thermal conductance is significantly reduced by 82% compared to $BN_1$. Putting $BN_5$ and $BN_1$ into relation, the reduction is around 24%, which is slightly larger than the reduction of 13% for the similar biphenyl molecules $BP_1$ and $BP_4$ of figure 8.

## S4. COMPARISON OF MECHANISMS TO SUPPRESS PHONON HEAT TRANSPORT

Finally, we quantify the size of the four presented mechanisms to suppress phononic energy transport. As listed at the beginning of section S3 the mechanisms are (i) acetylene terminal blocks, (ii) substituents, (iii) meta coupling, and (iv) molecule-internal torsion. We choose molecule C from figure 3 to study the influence of points (i) to (iv), since acetylene end-groups, substituents, a meta coupling and a finite twist are all present. We quantify the effects of all mechanisms by altering the structure and comparing it to the original configuration or the molecular backbone. The analysis is presented in figure S9, showing the different molecular structures in panel (a), cumulative thermal conductances in panel (b), and the relevant terminal propagator elements in panel (c). Please note that we use capital letters A to G to identify the molecules in the analysis. The labels should not be confused with those of figure 3 in the main text.

To study the first mechanism of mode filtering by terminal building blocks, we remove the acetylene linkers, see figure S9(a), and compare the transport properties of B to the original molecule A, generated by the genetic algorithm. The cumulative thermal conductance and especially the saturation values show a significant change due to acetylene terminal building blocks. Obtaining $\kappa_{ph}(600\ \mathrm{K}) = 0.35$ pW/K for A and $\kappa_{ph}(600\ \mathrm{K}) = 5.11$ pW/K for B, the thermal conductance is suppressed by 93% due to the presence of the acetylenes. The large difference in the phononic thermal conductances between structures A and B arises mainly from the energy ranges 3 to 6 meV and 8 and 10 meV, see figure S9(b). In these two energy intervals the contributions of all the modes of $xx$, $yy$ and $zz$ character are enhanced for B, see figure S9(c). In addition, resonances in the propagator in figure S9(c) are broader for B than for A. As a word of caution we note that due to the finite twist angle and the tilted molecular geometries, the Cartesian components of the propagators of A and B in figure S9(d) can only be interpreted approximately as transversal or longitudinal since modes mix.

Substituents can induce mass disorder, related destructive interferences in the phononic transmission and finite twist angles due to steric repulsion between building blocks. In order to quantify their influence, we compare molecule A to the backbone C. Due to the absence of substituents in C, the twist angle is reduced, and mass disorder or related destructive interferences disappear. The thermal conductance of A is clearly smaller than those of C. According to the cumulative thermal conductance, these differences mainly arise from around 6 to 7.5 meV and 9 to 11 meV. The propagator elements reveal that longitudinal $xx$ modes as well as those of kind $yy$ are strongly enhanced for C in these energy regions. The thermal conductance of the molecular backbone C is already quite low because the genetic algorithm also optimizes the selection of the building blocks.

We want to identify now, which substituents have the largest influence on $\kappa_{ph}(T)$. For this purpose, the masses of the three substituents in the middle of molecule A are set to the mass of hydrogen in configuration D. (The atomic structure of molecule D resembles those of molecule G but retains the twist angle of $87.7°$ of molecule A.) Given $\kappa_{ph}(600\ \mathrm{K}) = 0.35$ pW/K for A and $\kappa_{ph}(600\ \mathrm{K}) = 2.04$ pW/K for D, the thermal conductance is suppressed by 82.8% for A. The most important energy ranges leading to significantly different cumulative thermal conductance contributions in figure S9(b) are 7.5 to 9 meV and 10 to 12 meV. Structure A shows antiresonances especially for the $xx$ component of the propagator in these ranges, whereas structure D features particularly large contributions of precisely these longitudinal parts. Therefore, the bromine substituents in vicinity of the anthracene-anthracene connection are particularly important to suppress transmission maxima and thus the thermal conductance of structure A. As a cross-check, we remove all those substituents in configuration E, which have no influence on the twist of molecule A. The thermal conductance of E is only slightly altered as compared to A, setting it into relation with the suppression observed for D. The effect of the substituents, which are not responsible for twist, is hence rather negligible. To conclude, we assign a greater importance to those substituents that cause the twist of building blocks. Obviously destructive interferences are particularly effective in reducing vibrational heat transport if transmission maxima of the unsubstituted molecule are suppressed. Thus, the genetic algorithm implicitly optimizes energies of destructive interferences to be located at transmission maxima, when attachment positions of substituents are searched.

Next, we investigate the effect of meta- versus para-configurations. To do so, we compare structure A, where the anthracene rings are connected in a meta configuration, to structure F, where the coupling is replaced by a para-linkage. The thermal conductance of structure F is almost a factor of 2.5 larger than those of A. According to the cumulative thermal conductance, this difference is mainly caused by phonon modes between 3 and 4 meV as well as 8 and 10 meV. The propagator elements reveal that structure F exhibits increased vibrational contributions mainly of types $xx$ and $yy$ in these energy ranges.

Finally, we study the effect of the twist angle by comparing configurations D and G. As explained above, the masses of the three substituents in the middle of the molecule are set to those of hydrogen in configuration D, while preserving the geometry of molecule A. These three substituents are also replaced by hydrogens in configuration G, but the geometry is optimized subsequently. Structures D and G are thus identical except for the twist angle, which amounts to $87.7°$ and $37.9°$, respectively.

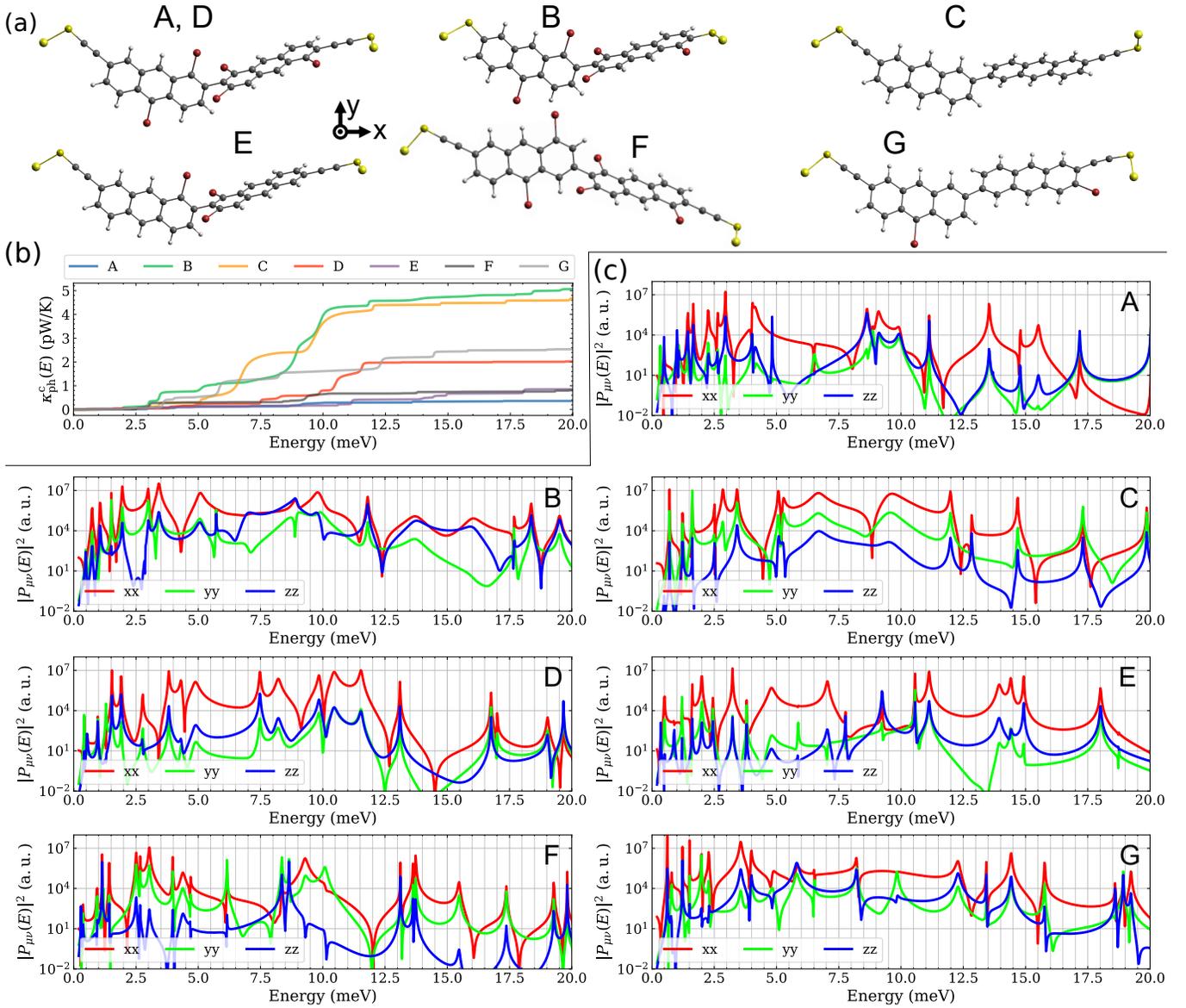

FIG. S9: (a) Molecular structures based on molecule C of figure 3. Different configurations are analyzed with the following properties. A: original structure, B: acetylene end groups removed, C: molecular backbone without any substituents, D: original structure with mass manipulated bromine atoms near the anthracene-anthracene connection (responsible for the molecule-internal twist), E: substituents not responsible for the twist angle removed, F: para configuration of the original structure, G: substituents responsible for the tilt angle replaced by hydrogens (identical to structure D except for the twist angle). (b) Transmission of all molecules as a function of energy. (c) Cumulative thermal conductance as a function of energy at 300 K. The saturation values of the thermal conductance at 600 K (not shown) are A: $\kappa_{ph} = 0.35$ pW/K, B: $\kappa_{ph} = 5.11$ pW/K, C: $\kappa_{ph} = 4.71$ pW/K, D: $\kappa_{ph} = 2.04$ pW/K, E: $\kappa_{ph} = 0.87$ pW/K, F: $\kappa_{ph} = 0.81$ pW/K and G: $\kappa_{ph} = 2.56$ pW/K. (c) Absolute square of terminal propagator elements from left to right gold anchor atoms as a function of energy for the studied molecules.

In accordance with our previous discussion, the thermal conductance of D is lowered by 20% compared to G due to the increased twist angle. The final rise of the thermal conductance of G beyond those of D results from strong suppressions of both $xx$ and $yy$ components near 12.5 and 14.5 meV for D, whereas G shows maxima. In summary, the reduction of the thermal conductance for increased twist angles is found for this molecule as well. Since transport is carried to a great extent by longitudinal modes, the influence of the twist angle is smaller compared to the analysis in figure 8.

In summary, all four mechanisms (i) to (iv) have a decisive influence on the phononic transport properties. Most important for the studied molecule are the endgroups (mechanism 1), followed by mass disorder and destructive interferences from substituents

(mechanism 2), meta vs. para coupling (mechanism 3) and finally the molecule-internal twist (mechanism 4).

---


\fi

\end{document}